\documentclass[prl,aps,twocolumn,showpacs,floatfix,amsmath,amssymb,aps,longbibliography]{revtex4-2}
\usepackage{amsfonts}
\usepackage{amssymb}
\usepackage{tikz-cd}
\usepackage[linesnumbered,ruled,vlined]{algorithm2e}
\usepackage{hyperref}
\usepackage{graphicx}
\usepackage{dcolumn}
\usepackage{bm,amsmath,verbatim}
\usepackage{mathrsfs}
\usepackage{color}
\usepackage{mathtools}
\usepackage{amsthm}
\newtheorem{theorem}{Theorem}

\newtheorem{lemma}[theorem]{Lemma}
\theoremstyle{definition}
\newtheorem{definition}[theorem]{Definition}
\theoremstyle{plain}
\usepackage[T1]{fontenc}
\usepackage[utf8]{inputenc}
\usepackage{booktabs}
\usepackage{amstext}
\usepackage{amsopn}
\usepackage{algorithmicx}
\usepackage{algpseudocode}
\hypersetup{colorlinks,linkcolor=blue,citecolor=blue,filecolor=blue,urlcolor=blue}

\newcommand{\HideTOCEntries}{%
  \let\SavedContentsLine\contentsline
  \renewcommand{\contentsline}[4]{}%
}
\newcommand{\ShowTOCEntries}{%
  \let\contentsline\SavedContentsLine
}

\begin{document}
\title{Universally Certifying Partially Self-Correcting Quantum Memories in 3D Translation-Invariant Stabilizer Codes}
\author{Chong-Yuan Xu$^{1}$}
\email{xu-cy24@mails.tsinghua.edu.cn}
\author{Yong Xu$^{1,2}$}
\email{yongxuphy@tsinghua.edu.cn}
\affiliation{$^{1}$Center for Quantum Information, IIIS, Tsinghua University, Beijing 100084, People's Republic of China}
\affiliation{$^{2}$Hefei National Laboratory, Hefei 230088, People's Republic of China}

\begin{abstract}
Achieving passive protection of quantum information at finite temperature is a central problem at the intersection of quantum information and quantum many-body physics.   
Yet a general, computable criterion for determining self-correcting behavior directly from the form of local interactions remains elusive. 
Here we develop a unified algebraic-geometric framework for certifying partial self-correction in three-dimensional translation-invariant local stabilizer codes, encompassing the cubic code as a special case.
From the algebraic curves defined by the codes’ defining polynomials, we construct valuation matrices on the boundaries of their normalized projective closures and derive a computable rank criterion. 
Together with a polynomial-time decoder, this criterion guarantees super-Arrhenius memory lifetimes for code families satisfying it under the specified local Davies dynamics. 
We apply the framework to certify several classes of partially self-correcting quantum memories, including nearest-neighbor models with stabilizer weight as low as six and long memory lifetimes at modest system sizes. 
These results establish an algebraic-geometric connection between local stabilizer structure and the recoverability of quantum information at finite temperature, providing a systematic approach to certifying and designing passive quantum memories.
\end{abstract}
\maketitle


How to stably store quantum information in quantum systems with local interactions in three or fewer spatial dimensions is a fundamental question shared by quantum information, quantum computing, and quantum many-body physics~\cite{dennis2002topological,hamma2009toric,chesi2010self,yoshida2011feasibility,pastawski2011quantum,wootton2012quantum,hastings2013self,wootton2013topological,bombin2013self,brown2014entropic,hutter2014two,bahri2015hot,vijay2015new,pastawski2015fault,roberts2020symmetry,stahl2023self,metz2023self}.
Self-correcting quantum memories (SCQMs) use a system's intrinsic interactions to suppress logical errors induced by thermal noise and are classified as fully self-correcting quantum memories (FSCQMs) or partially self-correcting quantum memories (PSCQMs) according to the dependence of their lifetimes on system size and temperature~\cite{terhal2015quantum,brown2016quantum,siva2017topological}.
A PSCQM extends its memory lifetime by increasing system size within a temperature-dependent finite-size window and achieves super-Arrhenius lifetime scaling at low temperatures when the size is chosen appropriately~\cite{bravyi2013quantum}.
Identifying PSCQMs can both reveal the physical mechanisms by which local many-body systems protect quantum information and provide practical passive protection for finite-size devices, reducing the overhead of repeated measurements in active error correction~\cite{dennis2002topological,bacon2006operator,hutter2012self}.

Constructing SCQMs from quantum error-correcting codes is a promising yet challenging route~\cite{kay2008quantum,chesi2010thermodynamic}.
Two-dimensional local stabilizer codes, exemplified by toric codes, have been rigorously proved incapable of realizing SCQMs~\cite{kitaev2003fault,castelnovo2007entanglement,nussinov2008autocorrelations,bravyi2009no,alicki2009thermalization,bravyi2010tradeoffs,landoncardinal2013local}.
Higher spatial dimensions or long-range couplings offer established routes around this no-go theorem, with many encouraging advances reported in recent years~\cite{castelnovo2008topological,alicki2010thermal,michnicki2014three,brell2016proposal,stahl2021symmetry,williamson2024layer,guo2026toward,gu2026layer,williamson2026partial}.
The cubic code is an important representative of PSCQMs: it can sustain long memory lifetimes at low temperatures in finite-size systems~\cite{haah2011local,bravyi2011energy,bravyi2013quantum,williamson2016fractal}.
Subsequent studies have shown that the cubic code and a range of three-dimensional (3D) translation-invariant stabilizer codes can be equivalently represented as trivariate bicycle (TB) codes, ensuring their regular and simple local structure~\cite{haah2013commuting,bravyi2024high,voss2025multivariate,shaw2025lowering,berthusen2025toward}.
This broad space of models raises two closely related questions.
First, can we identify additional TB code families with superior performance that realize PSCQMs?
Second, can we determine directly from the generating polynomials whether a TB code family realizes a PSCQM?
Although general sufficient conditions for logarithmic energy barriers are known~\cite{bravyi2011energy,bravyi2013quantum,haah2013commuting}, 
verifying the required topological-order and no-strings conditions for a given code can require substantial 
model-specific analysis~\cite{bravyi2011energy,bravyi2013quantum,haah2013commuting,yoshida2013exotic}. 
This motivates us to develop an explicit criterion for partial self-correction in 3D
translation-invariant stabilizer codes that can be evaluated directly from the generating polynomials.

Here we establish a unified algebraic-geometric criterion for a logarithmic lower bound on logical energy barriers of TB code families. 
Combined with efficient recovery, this bound guarantees that families satisfying the criterion constitute PSCQMs under local Davies dynamics~\cite{davies1973harmonic,davies1974markovian}. 
Starting from the one-dimensional algebraic curves defined by the generating polynomials, we characterize the constraints on thermal-excitation transport that govern self correction through the orders of zeros and poles of coordinate functions at boundary points of the normalized projective closure of each irreducible branch, thereby reducing this energy-barrier condition to algebraic tests that can be performed directly using computer algebra. 
Using this criterion, we discover a series of novel code families forming PSCQMs that maintain the advantages of the cubic code's regular local structure and whose representative instances exhibit long memory lifetimes at finite sizes. In particular, the nearest-neighbor models achieve stabilizer weights as low as six, lower than that of the cubic code, further simplifying the implementation. These results reveal the algebraic-geometric structure underlying quantum self correction, establish a computable theoretical foundation for systematically designing 3D translation-invariant PSCQMs, and provide concrete candidate models with simple structures and excellent performance.


\textit{TB codes}---We first briefly review the construction 
of TB codes based on Laurent-polynomial representation~\cite{voss2025multivariate,xu2026topological}.
Let 
$
\mathcal R=\mathbb F_2[x^{\pm1},y^{\pm1},z^{\pm1}]
$
be the Laurent polynomial ring over the binary field $\mathbb F_2$. A TB code is specified 
by two polynomials $f,g\in \mathcal R$. The X- and Z-type stabilizer generators associated
with a monomial $r\in \mathcal M \subset \mathcal R$, which labels a cell containing a left and right data 
qubit, are represented by $(fr \mid gr)$ and $(g^Tr \mid f^Tr)$, respectively, 
where $f^T=f(x^{-1},y^{-1},z^{-1})$ and similarly for $g^T$.
Up to an overall phase, pure X- and Z-type Pauli strings $ E_X$ and $ E_Z$ are 
represented by pairs $(\varepsilon_1^X\mid\varepsilon_2^X)$ and $(\varepsilon_1^Z\mid\varepsilon_2^Z)$, 
respectively, with $\varepsilon_1^X,\varepsilon_2^X,\varepsilon_1^Z,\varepsilon_2^Z \in \mathcal R$.
The nonzero monomials in the first and second entries record the cells 
whose left and right data qubits are acted on.
The Pauli string $ E_X$ ($ E_Z$) generates syndromes of the $Z$ ($X$) checks, described 
by the syndrome polynomials 
$
\operatorname{syn}_Z( E_X)=g\varepsilon_1^X+f\varepsilon_2^X
$
[$
\operatorname{syn}_X( E_Z)=f^T\varepsilon_1^Z+g^T \varepsilon_2^Z
$].
Each nonzero syndrome monomial labels a violated check.

For a finite system with periodic boundary conditions, 
we work in the quotient ring
$
\mathcal R_L=\mathcal R/I_L,
$
where $I_L=\langle x^L+1,y^L+1,z^L+1\rangle$ is an ideal with $L$ being the system size, 
so that $x^L=y^L=z^L=1$. 
In the stabilizer generators, $f$ and $g$ are replaced by $\pi_L(f)$ and $\pi_L(g)$, respectively,
where $\pi_L:\mathcal R\to\mathcal R_L$ is the quotient map defined by $\pi_L(h)=h+I_L$,
and $r$ is chosen from monomials in $\mathcal R_L$.
Fixing $f,g$ and varying $L$ defines a TB code family with parameters $[[n_L,k_L,d_L]]$, 
where $n_L=2L^3$, and stabilizer group $\mathcal S_L$.
See more details in Supplemental Material Sec.~S-1.


\begin{figure}[t]
	\centering
	\includegraphics[width=\columnwidth]{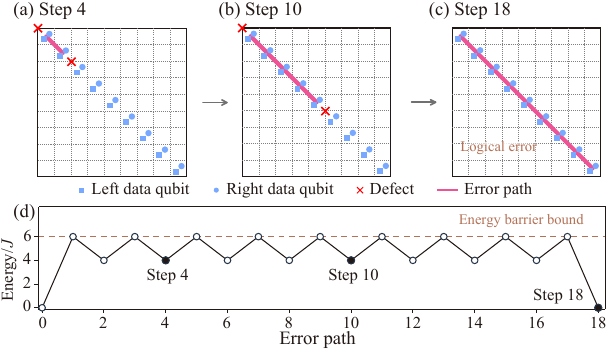}
	\caption{
		A periodic logical string and its actual energy path in a rank-deficient TB-code instance. 
		We take $f=1+x+y$, $g=1+x+z$, and $L=9$. 
		For the unique minimal prime ideal $P=\langle f,g\rangle$, $\operatorname{rank}_{\mathbb Q}V_P=2$ and $\ker_{\mathbb Z}V_P=\mathbb Z(0,1,-1)$. 
		(a)--(c) All three snapshots show the $x=0$ slice of the 3D periodic lattice, with $y$ along the horizontal direction and $z$ along the vertical direction. 
		The kernel vector $(0,1,-1)$ specifies the diagonal extension direction within this slice. 
		The path successively visits $p_r=(0,1+r,1-r)$ for $r=0,\ldots,8$. 
		(d) The energy barrier of a Pauli operator is the minimum peak excitation energy above a ground state over all sequences of single-qubit Pauli operations implementing it from the identity. 
		Each violated stabilizer costs $2J$.
		The displayed path gives an energy barrier of at most $6J$ for this logical operator.
	}
	\label{fig:necessity-string}
\end{figure}

\textit{Criterion}---We now use the stabilizer generators $S^{X}_r$ and $S^{Z}_r$ to construct the Hamiltonian,
\begin{equation}
H=-J\sum_{r\in \mathcal{M}} \left(S^{X}_r+S^{Z}_r \right),
\end{equation}
where $J>0$.
We provide the following algebraic criterion: for fixed $f,g\in\mathcal R$ and growing $L$, they constitute PSCQMs if all of the following conditions hold:
\begin{equation}
\boxed{
\begin{aligned}
&\langle f,g\rangle\subsetneq\mathcal R,\quad \gcd(f,g)=1,\\
&\forall P: P\in\operatorname{Min}(\langle f,g\rangle)\implies\operatorname{rank}V_P=3.
\end{aligned}
}
\label{eq:algebraic-criterion}
\end{equation}
Here $\langle f,g\rangle$ denotes an ideal generated by $f,g$ in $\mathcal R$, and 
$\gcd(f,g)=1$ means that their common divisors are units, namely Laurent monomials~\cite{dummit2004abstract,weibel2013kbook}.
The set $\operatorname{Min}(\langle f,g\rangle)$ consists of minimal prime ideals 
containing $\langle f,g\rangle$~\cite{atiyah2018introduction}.

Defining $V_P$ requires the language of algebraic geometry.
Consider a $P\in\operatorname{Min}(\langle f,g\rangle)$.
As $P$ is prime, $\mathcal R/P$ is an integral domain and determines an affine scheme $\operatorname{Spec}(\mathcal R/P)$~\cite{atiyah2018introduction}.
Under the first two conditions in Eq.~\eqref{eq:algebraic-criterion}, this scheme is a one-dimensional integral affine curve~\cite{hartshorne2013algebraic,js2022abstract}.
We construct its normal projective model in two steps.
First, choose a graded coordinate ring $\mathcal C_P$ such that $\operatorname{Proj}(\mathcal C_P)$ is a projective closure of $\operatorname{Spec}(\mathcal R/P)$.
This is an integral projective curve containing $\operatorname{Spec}(\mathcal R/P)$ as a dense open subset by immersion map $\nu_{P,1}$ with finite complement; both curves share the same function field~\cite{hartshorne2013algebraic,goertz2020algebraic}.
Next, take the integral closure $\overline{\mathcal C}_P$ of $\mathcal C_P$ in its ordinary fraction field.
The curve $\operatorname{Proj}(\overline{\mathcal C}_P)$ is the normalization of $\operatorname{Proj}(\mathcal C_P)$.
The normalization map $\nu_{P,2}$ replaces each of the finitely many singular points by one or more normal points, 
and is an isomorphism over the normal locus, and preserves the function field~\cite{hartshorne2013algebraic,goertz2020algebraic,silverman2009arithmetic}.

The quotient map and natural fraction-field embedding send $x,y,z\in\mathcal R$ to rational functions on $\operatorname{Spec}(\mathcal R/P)$.
The two function-field identifications above carry them to nonzero rational functions $\bar x,\bar y,\bar z$ on $\operatorname{Proj}(\overline{\mathcal C}_P)$.
These functions are invertible on the full inverse image of the affine curve.
Let $S_P=\{Q_1,\ldots,Q_{|S_P|}\}$ be the finite set of closed points complementary to $\nu_{P,2}^{-1}\circ\nu_{P,1}(\operatorname{Spec}(\mathcal R/P))$ in $\operatorname{Proj}(\overline{\mathcal C}_P)$.
Thus all zeros and poles of $\bar x,\bar y,\bar z$ lie in $S_P$.
For $h\in\{\bar x,\bar y,\bar z\}$, the integer $\operatorname{ord}_{Q_i}(h)$ records its signed order at $Q_i$: a zero of order $n$ gives $n$, a pole of order $n$ gives $-n$, and a regular, nonvanishing function gives $0$~\cite{atiyah2018introduction,hartshorne2013algebraic}.
These orders define
\begin{equation}
V_P=
\begin{pmatrix}
\operatorname{ord}_{Q_1}(\bar x)&\operatorname{ord}_{Q_1}(\bar y)&\operatorname{ord}_{Q_1}(\bar z)\\
\vdots&\vdots&\vdots\\
\operatorname{ord}_{Q_{|S_P|}}(\bar x)&\operatorname{ord}_{Q_{|S_P|}}(\bar y)&\operatorname{ord}_{Q_{|S_P|}}(\bar z)
\end{pmatrix}.
\label{eq:criterion-VP}
\end{equation}
The rank of this $|S_P|\times3$ integer matrix is taken over $\mathbb Q$, and is denoted by $\operatorname{rank}V_P$.

Although $\mathcal C_P$ is not unique, its choice does not affect $\operatorname{rank}V_P$; 
Supplemental Material Sec.~S-3 gives a conventional construction.
The gcd, projective closure, and normalization can be computed using the 
standard computer-algebra tools~\cite{developer2019sagemath,decker2024singular}.
Further details of the criterion are also provided in Supplemental Material Sec.~S-3.


We now give a brief proof for the criterion, deferring the complete proof to
Supplemental Material. 
The key concepts connecting TB codes to quantum self-correction are 
the topological-order condition (TOC) and the uniform no-strings condition (UNSC)~\cite{bravyi2011energy}.
Both have formulations for finite TB code families and for the infinite TB code obtained as $L\to\infty$.

\begin{figure}[t]
	\centering
	\includegraphics[width=\columnwidth]{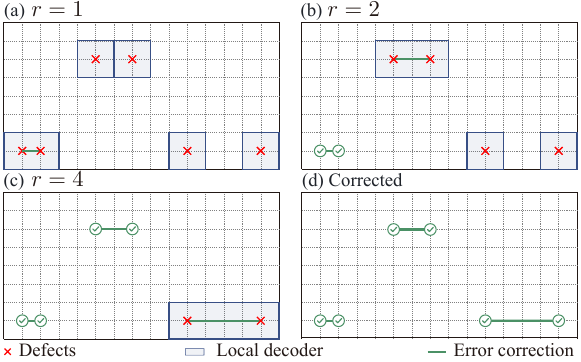}
	\caption{
		Two-dimensional illustration of RG decoding on a toric-code patch. 
		Red crosses denote defects, blue boxes local decoding regions, and green lines correction operators. 
		(a) At $r=1$, the adjacent lower-left defects are annihilated. 
		(b) At $r=2$, the upper defects merge into a cluster and are annihilated. 
		(c) At $r=4$, the final two defects are annihilated. 
		(d) The decoder returns the product of the recorded corrections. 
		For TB codes, each local correction is found from the actual parity-check matrices.
	}
	\label{fig:rg_2d_decoder}
\end{figure}

For an infinite code, UNSC states that there is a constant $\alpha>0$ such that, 
whenever the entire syndrome of a finite-support error lies in two disjoint cubes 
of side length $2\rho+1$ and neither cluster can be created separately by a 
finite-support error, their centers have $\ell_\infty$ separation at most $\alpha\rho$.
Here, the $\ell_\infty$ separation between two points refers to the largest 
coordinate difference.
Such clusters are called charged.
We express this condition as an algebraic one.
To demonstrate this, we consider an $X$-type Pauli error
$(\varepsilon_1\mid\varepsilon_2)$, whose syndrome is
\begin{equation}
g\varepsilon_1+f\varepsilon_2=r_1s_1+r_2s_2,
\label{eq:main-two-cluster-syndrome}
\end{equation}
where $r_1,r_2$ are monomials locating the two cube centers, respectively,
and $s_1,s_2\in \mathcal R$.
For a nonzero Laurent polynomial $h$, define $\operatorname{Deg}(h)$ as the maximum absolute coordinate exponent among its monomials.
The cube sizes give $\operatorname{Deg}(s_i)\le\rho$, and the charged-cluster assumption gives $r_1s_1,r_2s_2\notin\langle f,g\rangle$.
Setting $r=r_1^{-1}r_2$, the center separation is precisely $\operatorname{Deg}(r)$.

Previous work shows that local topological order and UNSC give 3D translation-invariant stabilizer-code families a logical energy barrier of $\Omega(\log L)$~\cite{bravyi2011energy}.
Combined with thermal recovery estimates, a polynomial-time decoder correcting all errors below a logarithmic energy-barrier threshold then yields PSCQM under the local Pauli Davies dynamics~\cite{bravyi2013quantum,williamson2026partial,gu2026layer}.
Our key observation is that TOC and UNSC for the infinite TB code imply local topological order and UNSC for the finite periodic family at a scale proportional to $L$.
For $\langle f,g\rangle\subsetneq\mathcal R$, our previous work establishes that infinite-code TOC is equivalent to $\gcd(f,g)=1$~\cite{xu2026topological}.
It therefore remains to derive UNSC from the criterion.
We prove that, on a suitable irreducible component $P$, the endpoint relation bounds every boundary order by $|\operatorname{ord}_Q(\bar r)|\le C\rho$ for $Q\in S_P$, with $C$ depending only on $f,g$.
For $r=x^ay^bz^c$, additivity gives $\operatorname{ord}_Q(\bar r)=a\operatorname{ord}_Q(\bar x)+b\operatorname{ord}_Q(\bar y)+c\operatorname{ord}_Q(\bar z)$, or, in matrix form,
\begin{equation}
V_P(a,b,c)^T=\bigl(\operatorname{ord}_Q(\bar r)\bigr)_{Q\in S_P}.
\label{eq:main-translation-orders}
\end{equation}
Full column rank of $V_P$ therefore bounds the displacement: $\operatorname{Deg}(r)=\max\{|a|,|b|,|c|\}\le\alpha\rho$.
Since every component has full rank, a uniform $\alpha$ can be chosen over the finitely many components.
Combined with the efficient decoder described below, this proves sufficiency.
The detailed proofs are given in Supplemental Material Sec.~S-4.

\begin{table}[t]
	\centering
	\caption{
		More weight-six TB code families that satisfy the criterion, thus realizing PSCQMs.
		$|S_P|$ is the total number of closed points on the boundary.
		$M$ is a submatrix consisting of three rows of $V_P$, with its determinant evaluated over the integers.
		$|\det M|>0$ ensures that the valuation matrix has full rank.
	}
	\label{tab:app_more_families}
	\setlength{\tabcolsep}{2pt}
	\renewcommand{\arraystretch}{1.3}
	\begin{ruledtabular}
		\begin{tabular}{cccc}
			$f$ & $g$ & $|S_P|$ & $|\det M|$ \\
			\midrule
			$1+x+y$ & $1+xy+z$ & 4 & 1 \\
			$1+x+y$ & $x+z+z^{-1}$ & 4 & 2 \\
			$x+x^{-1}+y$ & $y+y^{-1}+z$ & 4 & 4 \\
			$x+x^{-1}+y$ & $z+z^{-1}+y^{-1}$ & 4 & 16 \\
			\addlinespace
			$1+x+y$ & $1+xy^{2}+z$ &4 & 1 \\
			$1+x+y$ & $1+x^{2}y+z$ & 4 & 1 \\
			$1+x+y$ & $1+xy^{4}+z$ & 5 & 1 \\
			\addlinespace
			$1+x^{2}+y$ & $1+xy+z$ &  4  & 2\\
			$1+x^{2}+y$ & $1+xy^{2}+z$ &  5 & 2 \\
			$1+x^{2}+y$ & $1+xy^{3}+z$ &4  & 2 \\
			\addlinespace
			$1+x^{3}+y$ & $1+xy^{2}+z$ &5 & 1 \\
			$1+x^{3}+y$ & $1+x^{2}y+z$ &  5  & 1 \\
			$1+x^{3}+y$ & $1+xy^{4}+z$ &  6 & 1 \\
			\addlinespace
			$1+x^{4}+y$ & $1+xy+z$ & 5  & 4 \\
			$1+x^{4}+y$ & $1+xy^{2}+z$  & 4  & 4 \\
			$1+x^{4}+y$ & $1+xy^{3}+z$ & 5 & 4  \\
		\end{tabular}
	\end{ruledtabular}
\end{table}

\begin{figure*}[htbp]
	\centering
	\includegraphics[width=\textwidth]{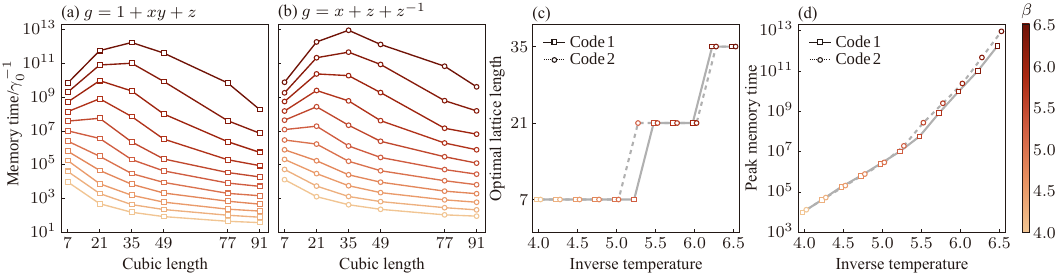}
	\caption{
		Numerical memory lifetimes of two TB code families under a Pauli Davies bath with Glauber transition rates, using BP readout with fixed decoder parameters.
		(a), (b) Lifetime versus cubic size $L$ for $g=1+xy+z$ and $g=x+z+z^{-1}$, respectively, with $f=1+x+y$ in both cases; color indicates the inverse temperature $\beta=4,4.25,\ldots,6.5$.
		Time is measured in units of $\gamma_0^{-1}$.
		Each point has 400 planned independent trajectories.
		Complete samples give the mean first detected readout-failure time, with error bars of one standard error; incomplete samples give lower bounds on the eventual mean of the same planned sample, as detailed in Supplemental Material Sec.~S-7.
		(c) Size at the largest plotted lifetime versus $\beta$.
		(d) Peak memory time versus $\beta$.
		For temperature groups containing incomplete samples, the peak size is provisional and the peak lifetime is a sample lower bound.
	}
	\label{fig:numerical_memory_time}
\end{figure*}

Our criterion is necessary and sufficient for a logarithmic lower bound on the logical energy barrier of the code family. 
Having established sufficiency above, we now turn to necessity.
If $\langle f,g\rangle=\mathcal R$, then no logical states exist; 
if $\gcd(f,g)\neq 1$, then the TOC fails, admitting a finite-support logical error~\cite{xu2026topological}.
It therefore remains only to show that rank deficiency permits a logical error channel with bounded energy barrier.
If $\operatorname{rank}V_P<3$, then there exists a nonzero integer vector $\vec v=(a,b,c)^T\in\ker_{\mathbb Z}V_P$.
We set $m=x^ay^bz^c$.
Since $V_P\vec v=0$, its image $\bar m$ has zero order at every boundary point and hence is a nonzero element of the curve's finite constant field.
Its multiplicative order $q$ gives $1+m^q\in P$.
Nilpotence in the local quotient ring and clearing localization denominators yield fixed $s\ge0$ and $h\notin P$ such that $h(1+m^q)^{2^s}\in\langle f,g\rangle$.
In characteristic $2$, this is $h(1+t)\in\langle f,g\rangle$, with $t=m^{q2^s}$.
Both $h$ and $ht$ are charged, and a finite-support $X$ operator $(\varepsilon_1\mid\varepsilon_2)$ has syndrome $g\varepsilon_1+f\varepsilon_2=h+ht$.
Extending the string at one end gives
\begin{equation}
E_M=(1+t+\cdots+t^{M-1})(\varepsilon_1\mid\varepsilon_2),
\label{eq:main-string-growth}
\end{equation}
with syndrome $\operatorname{syn}(E_M)=h(1+t^M)$.
The two endpoints remain charged.
We show that, for infinitely many finite periods, the segment in Eq.~\eqref{eq:main-string-growth} closes into a nontrivial logical operator with an energy barrier bounded independently of system size.
Rank deficiency therefore produces constant-barrier logical errors depending only on $h$ and $t$ for infinitely many finite periods.
The full proof is given in Supplemental Material Sec.~S-4 D.

The factor $1+t$ thus emerges naturally from rank deficiency, giving a polynomial realization of the two-endpoint string segments considered by Ref.~\cite{bravyi2011energy,bravyi2013quantum} and directly specifying their extension direction.
Figure~\ref{fig:necessity-string} shows a finite periodic example: for $f=1+x+y$ and $g=1+x+z$, take $t=yz^{-1}$, $h=z$, and $(\varepsilon_1\mid\varepsilon_2)=(1\mid1)$.
The string extends along $(0,1,-1)$ and closes into a logical operator at $L=9$, with energy barrier at most $6J$.


\textit{Efficient decoders}---To recover the stored information efficiently, we specialize the renormalization group (RG) decoder to TB codes~\cite{bravyi2013quantum},
treating the X- and Z-type syndromes independently.
Beginning at scale $r=1$, we connect syndrome defects separated by an $\ell_\infty$ distance of at most $r$ and 
identify the resulting connected components as clusters.
For each cluster, we expand its bounding box by a fixed-width neighborhood and seek a correction supported within 
this region by solving the binary syndrome equations associated with the appropriate parity-check matrix $H_X$ or $H_Z$.
Each successful correction is recorded, and the corresponding defects are removed.
Unresolved clusters are carried forward to the next scale, with $r\leftarrow 2r$. Once the syndrome is empty, the 
decoder returns the accumulated Pauli correction. Figure~\ref{fig:rg_2d_decoder} illustrates clustering and local 
correction at scales $r=1,2,4$ in two dimensions; for TB codes, the local solver uses the corresponding TB parity-check matrices.

For code families satisfying Eq.~\eqref{eq:algebraic-criterion}, this decoder corrects every Pauli error with energy barrier below $a\log n_L+b$, for constants $a>0$ and $b$ independent of $L$, at all sufficiently large sizes.
With $O(\log n_L)$ scales and binary Gaussian elimination for the local solver, its runtime is $O(n_L^4\log n_L)$.
Together with the thermal recovery estimate, this establishes efficient recovery and the PSCQM guarantee.
The construction and correctability argument are detailed in Supplemental Material Sec.~S-6.


\textit{Results}---We prove that to satisfy Eq.~\eqref{eq:algebraic-criterion}, the 
partially self-correcting TB code families must have stabilizer weight of at least 6
(see Supplemental Material Sec.~S-5). 
We then identify infinitely many simple partially self-correcting TB code families.
In particular, 
we find some nearest-neighbor models with stabilizer weight of 6 as shown in 
Table~\ref{tab:app_more_families}.
In addition, consider the code family generated by
\begin{equation}
f=1+x+y,\quad g_a=1+x^a+z,\quad a\in\mathbb N^+.
\label{eq:numerical-weight-six-families}
\end{equation}
For every $a$, these polynomials are coprime and generate a proper prime ideal.
We prove in Supplemental Material Sec.~S-5 that the criterion holds if and only if $a\notin\{2^s:s\ge0\}$.
The smallest admissible exponent is thus $a=3$.

We now numerically investigate the finite-temperature storage performance of the two TB code families 
with $f=1+x+y$ and $g=1+xy+z$ or $g=x+z+z^{-1}$ and estimate their memory lifetimes.
The simulations use independent single-qubit Pauli Davies channels with Glauber transition rates~\cite{glauber1963time,bravyi2013quantum} 
and a fixed belief-propagation (BP) decoder~\cite{yedidia2003understanding,roffe2020decoding,roffe2022ldpc} at readout.
We estimate the mean time to the first detected readout failure of the full logical register, measured
in units of the inverse rate scale $\gamma_0^{-1}$.
The simulation method and decoder settings are detailed in Supplemental Material Sec.~S-7.

Figure~\ref{fig:numerical_memory_time} shows the memory lifetimes of both code families across 
a range of temperatures and system sizes. At sufficiently low temperatures, the memory time 
increases substantially with system size before reaching a maximum at an optimal size $L^*$. 
As the temperature decreases, this optimum shifts from the smallest instances to intermediate sizes, 
while the observed lifetimes increase by many orders of magnitude, exceeding 
$10^{12}\gamma_0^{-1}$ at $\beta=6.5$. 
Over the simulated temperature range, the peak memory time $T^*$ exhibits 
superexponential growth with $\beta$, with numerical fits yielding
$T^*\gamma_0\approx \exp({1.55\beta^2-8.85\beta+20.15})$ for $g=1+xy+z$ and 
$T^*\gamma_0\approx\exp({1.73\beta^2-10.08\beta+22.38})$ for $g=x+z+z^{-1}$.
These finite-size results are consistent with the expected behavior of partial self-correction~\cite{bravyi2013quantum}.


We have developed an algebraic-geometric framework for certifying partial self-correction in TB codes and identified a large, previously unrecognized class of structurally simple partially self-correcting codes. These results establish a connection between algebraic geometry and quantum self-correction, providing a foundation for the systematic discovery of further partially self-correcting quantum memories. The methods developed here may also extend to systems without translation symmetry.

\begin{acknowledgments}
This work is supported by Innovation Program for Quantum Science and Technology (Grant No. 2021ZD0301604) and the National Natural Science Foundation of China (Grant No. 12474265 and No. 11974201).
\end{acknowledgments}

\begin{widetext}
\setcounter{equation}{0} \setcounter{figure}{0} \setcounter{table}{0} 
\renewcommand{\theequation}{S\arabic{equation}} 
\setcounter{secnumdepth}{2}
\renewcommand{\thesection}{S-\arabic{section}}
\renewcommand{\thesubsection}{\Alph{subsection}}
\makeatletter
\renewcommand{\p@subsection}{\thesection\space}
\renewcommand{\@seccntformat}[1]{\csname the#1\endcsname.\ }
\renewcommand{\@hangfrom@section}[3]{\@hangfrom{#1#2}#3}
\makeatother

\addtocontents{toc}{\protect\ShowTOCEntries}
\begingroup
\HideTOCEntries
\tableofcontents
\endgroup

\section{Trivariate bicycle codes}

\label{sec:tb_codes}

In this section, we introduce the trivariate bicycle (TB) codes studied in this work and the polynomial description of their stabilizer generators, Pauli errors, and syndromes~\cite{voss2025multivariate,xu2026topological}.
It provides the detailed construction and notation underlying the main-text description and the analysis that follows.

A TB code is specified by two polynomials and the system's period.
We begin with the trivariate Laurent polynomial ring
\begin{equation}
\mathcal R=\mathbb F_2[x^{\pm1},y^{\pm1},z^{\pm1}],
\end{equation}
where $\mathbb F_2$ is the binary field.
Elements of $\mathcal R$ may contain negative integer powers of the variables.
Fix $f,g\in\mathcal R$ and write
\begin{equation}
f=\sum_{\vec v\in\mathbb Z^3}\lambda_{\vec v}x^{v_x}y^{v_y}z^{v_z},\quad g=\sum_{\vec v\in\mathbb Z^3}\mu_{\vec v}x^{v_x}y^{v_y}z^{v_z},
\end{equation}
where $\vec v=(v_x,v_y,v_z)$ and $\lambda_{\vec v},\mu_{\vec v}\in\mathbb F_2$.
The antipode map reverses the sign of each monomial's exponent vector; applied to $f$ and $g$, it gives
\begin{equation}
f^T=\sum_{\vec v\in\mathbb Z^3}\lambda_{\vec v}x^{-v_x}y^{-v_y}z^{-v_z},\quad g^T=\sum_{\vec v\in\mathbb Z^3}\mu_{\vec v}x^{-v_x}y^{-v_y}z^{-v_z}.
\end{equation}
In particular, $(f^T)^T=f$ and $(g^T)^T=g$.

We next construct the check matrices of a finite system from these polynomials.
Consider a three-dimensional (3D) simple cubic lattice with periodic boundary conditions.
For simplicity, take the period in each direction to be $L\in\mathbb N^+$.
Define the cyclic-shift matrix over $\mathbb F_2$ by
\begin{equation}
(T_L)_{ij}=\delta_{(i+1)\bmod L,j},\quad 0\le i,j<L.
\end{equation}
Let $I_L$ be the $L\times L$ identity matrix.
The shift matrices for the three lattice directions are
\begin{equation}
T_{x,L}=T_L\otimes I_L\otimes I_L,\ T_{y,L}=I_L\otimes T_L\otimes I_L,\ T_{z,L}=I_L\otimes I_L\otimes T_L.
\end{equation}
Their rows and columns are indexed by $(a,b,c)\in\mathbb Z_L^3$.
Substituting $x\mapsto T_{x,L}$, $y\mapsto T_{y,L}$, and $z\mapsto T_{z,L}$ into the Laurent polynomials defines the $X$- and $Z$-type check matrices
\begin{equation}
\begin{aligned}
H_X&=\bigl(f(T_{x,L},T_{y,L},T_{z,L})\mid g(T_{x,L},T_{y,L},T_{z,L})\bigr),\\
H_Z&=\bigl(g^T(T_{x,L},T_{y,L},T_{z,L})\mid f^T(T_{x,L},T_{y,L},T_{z,L})\bigr).
\end{aligned}
\label{eq:tb_checks}
\end{equation}
Each check matrix has $L^3$ rows and $2L^3$ columns.
The columns in its left and right blocks label two groups of data qubits, denoted by left and right, respectively.
The nonzero entries of each row specify the support of the corresponding $X$- or $Z$-type stabilizer generator.
The shift matrices are invertible and mutually commuting, and satisfy $T_{\alpha,L}^T=T_{\alpha,L}^{-1}$ for $\alpha=x,y,z$.
Thus, evaluating $f^T$ or $g^T$ at these matrices gives the transpose of the block obtained from $f$ or $g$, respectively.
Together with arithmetic over $\mathbb F_2$, this implies $H_XH_Z^T=0$.
The check matrices therefore define a valid Calderbank-Shor-Steane (CSS) code: the TB code specified by $f,g,L$.

Keeping $f,g$ fixed and allowing $L$ to range over the positive integers produces a TB code family.
Let $\mathcal S_L$ denote the stabilizer group of the code at size $L$.
The code at size $L$ has parameters $[[n_L,k_L,d_L]]$, where $n_L=2L^3$, while $k_L$ and $d_L$ are the number of logical qubits and the code distance, respectively.
Here, $f,g$ fix the local check structure, whereas $L$ controls the system size.

Polynomials also provide a common representation of stabilizer generators, Pauli errors, and syndromes.
For a fixed $L$, introduce the quotient ring
\begin{equation}
\mathcal R_L=\mathcal R/\langle x^L+1,y^L+1,z^L+1\rangle.
\end{equation}
Let $\pi_L:\mathcal R\to\mathcal R_L$ be the quotient map, and write $\pi_L(h)$ for the residue class of $h\in\mathcal R$.
The variables of this finite polynomial ring are
\begin{equation}
x_0=\pi_L(x),\ y_0=\pi_L(y),\ z_0=\pi_L(z),
\end{equation}
which satisfy $x_0^L=y_0^L=z_0^L=1$ and thus encode the periodic boundary conditions.
Every element of $\mathcal R_L$ has a unique expansion over $\mathbb F_2$ in the monomial basis
\begin{equation}
M_L=\{x_0^ay_0^bz_0^c\mid a,b,c\in\mathbb{N},\ a,b,c<L\}.
\end{equation}
The image of $f\in\mathcal R$ is
\begin{equation}
\pi_L(f)=\sum_{\vec v\in\mathbb Z^3}\lambda_{\vec v} x_0^{v_x\bmod L}y_0^{v_y\bmod L}z_0^{v_z\bmod L}.
\end{equation}
Multiplication in $\mathcal R_L$ adds the exponents modulo $L$, while addition uses the field operations of $\mathbb F_2$.

We now use elements of $\mathcal R_L$ to represent stabilizer generators, Pauli errors, and syndromes.
We first describe Pauli operators on the data qubits.
The tensor-product structure of the shift matrices indexes the rows and columns of each block of the check matrices by ordered triples $(a,b,c)$, with $0\leq a,b,c<L$.
For each such triple, let $h=x_0^ay_0^bz_0^c\in M_L\subset\mathcal R_L$.
We label the corresponding qubits by $(\text{left},h)$ and $(\text{right},h)$, according to whether their columns belong to the left or right block.
Up to a global phase, every pure $X$-type or pure $Z$-type Pauli operator is in one-to-one correspondence with a pair in $\mathcal R_L^2$, denoted by $(p_1^X\mid p_2^X)$ or $(p_1^Z\mid p_2^Z)$, respectively.
Specifically, if $p_1=\sum_{h\in M_L}\gamma_h h$ and $p_2=\sum_{h\in M_L}\eta_h h$, the pair $(p_1\mid p_2)$ represents the $X$-type and $Z$-type operators
\begin{equation}
P_X=\prod_{h\in M_L}X_{\text{left},h}^{\gamma_h}X_{\text{right},h}^{\eta_h},\quad P_Z=\prod_{h\in M_L}Z_{\text{left},h}^{\gamma_h}Z_{\text{right},h}^{\eta_h},
\end{equation}
respectively.

The same monomials $h$, without a left or right qubit label, index the rows of $H_X$ and $H_Z$.
With the row convention specified by $T_L$, the corresponding stabilizer generators are represented by
\begin{equation}
S_h^X\longleftrightarrow(\pi_L(f)h\mid\pi_L(g)h),\quad S_h^Z\longleftrightarrow(\pi_L(g^T)h\mid\pi_L(f^T)h).
\end{equation}
Let $\mathcal G_L^X$ and $\mathcal G_L^Z$ be the respective sets of generators with their row labels, and write $\mathcal G_L=\mathcal G_L^X\sqcup\mathcal G_L^Z$.
These generators need not be independent.

Up to a global phase, a general Pauli error has a unique decomposition $E=E_XE_Z$ into pure $X$-type and $Z$-type operators.
It is therefore uniquely specified by the two polynomial pairs
\begin{equation}
E_X\longleftrightarrow(\varepsilon_1^X\mid\varepsilon_2^X),\quad E_Z\longleftrightarrow(\varepsilon_1^Z\mid\varepsilon_2^Z),
\end{equation}
where $\varepsilon_1^X,\varepsilon_2^X,\varepsilon_1^Z,\varepsilon_2^Z\in\mathcal R_L$.

These polynomials also determine the error syndrome, defined as
\begin{equation}
\operatorname{syn}(E)=\{S\in\mathcal G_L\mid[E,S]_+=0\}.
\end{equation}
When $E$ acts on a state in the code space, this set records precisely the checks whose eigenvalues change from $+1$ to $-1$.
We separate the syndrome according to the type of check that is flipped:
\begin{equation}
\operatorname{syn}_X(E)=\operatorname{syn}(E)\cap\mathcal G_L^X,\quad\operatorname{syn}_Z(E)=\operatorname{syn}(E)\cap\mathcal G_L^Z.
\end{equation}
The two syndrome components take values in the power sets of $\mathcal G_L^X$ and $\mathcal G_L^Z$, respectively.
Each power set is in one-to-one correspondence with $\mathcal R_L$: to represent a subset of generators, we sum the monomials $h\in M_L$ associated with their row labels.
Thus, the polynomial representations of the two components are
\begin{equation}
\operatorname{syn}_X(E)\longleftrightarrow\sum_{S_h^X\in\operatorname{syn}_X(E)}h,\quad
\operatorname{syn}_Z(E)\longleftrightarrow\sum_{S_h^Z\in\operatorname{syn}_Z(E)}h.
\end{equation}
Computing the overlap parity shows that the syndrome sets and their polynomial representations satisfy
\begin{equation}
\operatorname{syn}_X(E)\longleftrightarrow \pi_L(f^T)\varepsilon_1^Z+\pi_L(g^T)\varepsilon_2^Z,\quad
\operatorname{syn}_Z(E)\longleftrightarrow \pi_L(g)\varepsilon_1^X+\pi_L(f)\varepsilon_2^X.
\end{equation}

This polynomial representation has a direct geometric interpretation.
The monomial $h=x_0^ay_0^bz_0^c\in M_L$ corresponds to the site $(a,b,c)$ of the periodic cubic lattice.
At each site, place a unit cell containing a left data qubit and a right data qubit, and mark an $X$ checker and a $Z$ checker.
The two checkers in the cell labeled by $h$ correspond to $S_h^X$ and $S_h^Z$, respectively.
This correspondence also specifies the connections between checkers and data qubits.
For the $X$ checker in that cell, each nonzero monomial of $\pi_L(f)h$ identifies a cell whose left data qubit is acted on, and each nonzero monomial of $\pi_L(g)h$ identifies a cell whose right data qubit is acted on.
The corresponding stabilizer generator is the product of Pauli $X$ operators on these qubits.
The $Z$ checker is described in the same way, with its connections specified by $\pi_L(g^T)h$ and $\pi_L(f^T)h$.
Nonzero monomials in a syndrome polynomial identify the cells containing the checks flipped by the error.

Finally, we also consider the infinite-lattice system defined by the same $f,g$, corresponding to the limit $L\to\infty$.
Replacing $\mathcal R_L$ by $\mathcal R$ gives the same representation for finite-support Pauli operators, finite products of stabilizer generators, and their syndromes. 
In this setting, the syndrome maps $\mathcal R^2\to\mathcal R$ are given by $g\varepsilon_1^X+f\varepsilon_2^X$ and $f^T\varepsilon_1^Z+g^T\varepsilon_2^Z$.


\section{Quantum self-correction}
\label{sec:psc_codes}

Quantum self-correction concerns the passive protection of encoded quantum information against thermal noise.
In this section, we use the stabilizer-code notation introduced above to define thermal evolution, final readout, 
and memory time for a stabilizer code family, then distinguish self-correcting properties by the dependence 
of memory time on system size and temperature and by the computational efficiency of readout.

Consider a stabilizer code family indexed by system size $L$.
Throughout this section, dependence on $L$ and $\beta$ is implicit, and $n$ denotes the number of physical qubits.
The associated stabilizer Hamiltonian is defined as
\begin{equation}
H=-J\sum_{S\in\mathcal G}S,\quad J>0,
\label{eq:stabilizer_hamiltonian}
\end{equation}
where $\mathcal G$ is a set of stabilizer generators.
All generators commute, and every state in the code space has eigenvalue $+1$ under each generator.
Thus, $H$ is frustration-free, and its ground space is precisely the code space.
A change in a generator's eigenvalue from $+1$ to $-1$ contributes an excitation energy of $2J$.

Storage begins with an arbitrary initial state $\rho_0$ in the code space.
We set $k_{\mathrm B}=\hbar=1$ and couple the system to a thermal bath at temperature $1/\beta$.
During storage, the system evolves solely under the specified system--bath dynamics.
Writing $\mathcal L_t^{\mathrm{evo}}$ for the quantum channel describing evolution over a time $t$, the state becomes
\begin{equation}
\rho(t)=\mathcal L_t^{\mathrm{evo}}(\rho_0).
\end{equation}
At final readout, decoding aims to recover the initial encoded state as faithfully as possible.
For a stabilizer code, this can be implemented by syndrome measurements, followed by the computation and application of a Pauli correction determined by the measured syndrome.
The entire recovery procedure is represented by a completely positive, trace-preserving map $\mathcal L^{\mathrm{dec}}$, yielding
\begin{equation}
\rho_1=\bigl(\mathcal L^{\mathrm{dec}}\circ\mathcal L_t^{\mathrm{evo}}\bigr)(\rho_0).
\end{equation}
The recovery procedure must work for any encoded state and must not depend on the unknown initial logical information.

We quantify the recovery error by the worst-case trace-norm difference between the initial and recovered states:
\begin{equation}
\delta(t)=\sup_{\rho_0}\left\|\rho_1-\rho_0\right\|_1,
\end{equation}
where the supremum is over all density operators in the code space.
For a fixed error tolerance $0<\varepsilon^*<1$, the memory time is defined as
\begin{equation}
T^{\mathrm{mem}}=\sup\bigl\{T\geq0\mid\delta(t)<\varepsilon^*,\ 0\leq t<T\bigr\}.
\label{eq:memory_lifetime}
\end{equation}
The dependence of $T^{\mathrm{mem}}$ on $\varepsilon^*$ is also implicit.
The recovery error must therefore remain below the tolerance throughout the storage interval.
The memory time is determined jointly by the chosen dynamics and recovery procedure.

A fully self-correcting quantum memory (FSCQM) permits arbitrarily long reliable storage at a fixed low temperature by increasing the system size~\cite{dennis2002topological,bacon2006operator}.
Specifically, if there is an efficient decoder for the entire code family with runtime $\operatorname{poly}(n)$ such that, for every fixed error tolerance $0<\varepsilon^*<1$, there exists a finite critical inverse temperature $\beta_c>0$ for which every fixed $\beta>\beta_c$ satisfies
\begin{equation}
\lim_{L\to\infty}T^{\mathrm{mem}}=\infty,
\end{equation}
then the code family constitutes an FSCQM under the chosen dynamics.

A partially self-correcting quantum memory (PSCQM) satisfies a weaker lifetime requirement: the system size may be chosen as a function of temperature, and the resulting memory time must grow sufficiently rapidly as the temperature decreases in the low-temperature limit~\cite{bravyi2013quantum}.
Specifically, if there exists a decoder with runtime $\operatorname{poly}(n)$ such that, for every fixed error tolerance $\varepsilon^*$, one can choose a finite size $L=L^*(\beta,\varepsilon^*)$ for which
\begin{equation}
\lim_{\beta\to\infty}\frac{\log T^{\mathrm{mem}}}{\beta}=\infty,
\end{equation}
then the code family constitutes a PSCQM under the chosen dynamics.
This condition defines super-Arrhenius growth.
In typical models of partial self-correction that do not satisfy the FSCQM condition, increasing the system size at a fixed temperature may improve the memory time only over a limited range.
The memory time may then reach a maximum or saturate, and a finite optimal size may exist.

A partially self-correcting quantum system (PSCQS) satisfies the same lifetime requirement, with no restriction on the computational complexity of recovery.
It suffices that a recovery procedure exists such that, for every fixed error tolerance, a temperature-dependent choice of system size yields the super-Arrhenius growth described above.
Thus, a PSCQS characterizes how long encoded information can be preserved and recovered, whereas a PSCQM additionally requires efficient decoding.

Under these definitions, the three properties obey the inclusion relations
\begin{equation}
\mathrm{FSCQM}\Longrightarrow
\mathrm{PSCQM}\Longrightarrow
\mathrm{PSCQS}.
\end{equation}

All the three concepts are defined relative to a specified dynamical evolution.
Even for the same stabilizer Hamiltonian, different system--bath coupling operators and transition rates can change the pathways and timescales of information loss.
Self-correcting properties must therefore be stated together with the thermal dynamics under consideration.
Following Refs.~\cite{alicki2009thermalization,chesi2010self,chesi2010thermodynamic,bravyi2013quantum,gu2026layer}, we use a Davies bath with local Pauli coupling channels to describe the behavior of the encoded system in a typical thermal environment with weak local coupling~\cite{davies1973harmonic,davies1974markovian}.
This is a standard framework for studying quantum self-correction, capturing thermal excitation and relaxation when the coupling is weak and local and the bath correlation time is sufficiently short~\cite{alicki2009thermalization,bravyi2013quantum,gu2026layer}.
In the Davies weak-coupling limit, the dynamics is Markovian and is governed by a Lindblad master equation.
Specifically, we take independent single-qubit $X,Z$ coupling channels.
Each local Pauli operator $\sigma_q$ is resolved into spectral components $A_{\sigma q\omega}$ according to the system energy change $-\omega$.
Here, $\sigma\in\{X,Z\}$ and $q$ labels a physical qubit.
The corresponding master equation is
\begin{equation}
\frac{d\rho}{dt}=-i[H,\rho]+\sum_{\sigma,q,\omega}\gamma(\omega)\biggl(A_{\sigma q\omega}\rho A^{\dagger}_{\sigma q\omega}-\frac12 \{A^{\dagger}_{\sigma q\omega}A_{\sigma q\omega},\rho\}\biggr).
\end{equation}
Here, $\omega$ is the energy transferred from the system to the bath, so $\omega>0$ corresponds to an energy-lowering transition.
The transition rates $\gamma(\omega)$ satisfy detailed balance,
\begin{equation}
\gamma(-\omega)=e^{-\beta\omega}\gamma(\omega),
\end{equation}
which exponentially suppresses energy absorption from the bath relative to the reverse, energy-lowering transition.
We choose
\begin{equation}
\gamma(\omega)
=\frac{\gamma_0}{1+e^{-\beta\omega}},
\label{eq:davies_rate}
\end{equation}
where $\gamma_0>0$ is independent of system size and temperature.


\section{Criterion}
\label{sec:criterion}

In this section, we present the technical details on the algebraic criterion stated in the main text.
Given $f,g\in\mathcal R$, where $\mathcal R=\mathbb F_2[x^{\pm1},y^{\pm1},z^{\pm1}]$, the criterion in the main text states that the TB code family defined by $f,g$ constitutes a PSCQM if all of the following conditions on $f,g$ hold:
\begin{equation}
\boxed{
\langle f,g\rangle\ne\mathcal R,\ \gcd(f,g)=1,\ \forall P: P\in\operatorname{Min}(\langle f,g\rangle)\implies\operatorname{rank}V_P=3.
}
\label{eq:criterion}
\end{equation}
We now introduce the notation and definitions used in these conditions, one by one.
The notation $\langle f,g\rangle$ denotes the ideal of $\mathcal R$ generated by $f,g$.
The condition $\langle f,g\rangle\ne\mathcal R$ means that this ideal is proper, or equivalently, that $1\notin\langle f,g\rangle$~\cite{atiyah2018introduction}.
The notation $\gcd(f,g)$ denotes the greatest common divisor of $f,g$, and $\gcd(f,g)=1$ means that $f,g$ have only units as common divisors~\cite{dummit2004abstract}.
An element of $\mathcal R$ is a unit if and only if it is a monomial~\cite{weibel2013kbook}.
The notation $\operatorname{Min}(\langle f,g\rangle)$ denotes the set of all minimal prime ideals over $\langle f,g\rangle$.
An ideal $P$ of $\mathcal R$ is a minimal prime ideal over $\langle f,g\rangle$ if and only if $P$ is prime, $\langle f,g\rangle\subseteq P$, and there is no other prime ideal $P'$ such that $\langle f,g\rangle\subseteq P'\subsetneq P$~\cite{atiyah2018introduction}.

The third condition is formulated in the language of algebraic geometry, whose framework we introduce below.
Fix an arbitrary $P\in\operatorname{Min}(\langle f,g\rangle)$.
Since $P$ is prime, the quotient ring $\mathcal R/P$ is an integral domain~\cite{atiyah2018introduction}.
We denote the affine scheme with coordinate ring $\mathcal R/P$ by $\operatorname{Spec}(\mathcal R/P)$.
Its points correspond to the prime ideals of $\mathcal R/P$~\cite{hartshorne2013algebraic}.
We now describe its topology and its sheaves of regular and rational functions.
The topology on $\operatorname{Spec}(\mathcal R/P)$ is the Zariski topology.
For any ideal $I$ of $\mathcal R/P$, define the closed set
\begin{equation}
V_I(\mathcal R/P)=\{Q\in\operatorname{Spec}(\mathcal R/P):I\subseteq Q\}.
\end{equation}
As $I$ ranges over all ideals of $\mathcal R/P$, these sets give all the closed sets.
Correspondingly, for any $h\in\mathcal R/P$, define the basic open set
\begin{equation}
D_h(\mathcal R/P)=\{Q\in\operatorname{Spec}(\mathcal R/P):h\notin Q\}.
\end{equation}
These basic open sets form a basis for the Zariski topology~\cite{hartshorne2013algebraic}.
The sheaves of regular and rational functions on the affine scheme are denoted by $\mathcal O(\mathcal R/P)$ and $\mathcal K(\mathcal R/P)$, respectively.
They assign a ring of functions to each open set and a restriction map to each inclusion of open sets, and satisfy the sheaf axioms.
In particular, functions can be restricted to smaller open sets, and local functions that agree on overlaps can be glued together uniquely.
On a basic open set $D_h(\mathcal R/P)$, the ring of regular functions is
\begin{equation}
\mathcal O_h(\mathcal R/P)=(\mathcal R/P)[h^{-1}],
\end{equation}
the localization obtained by making $h$ invertible~\cite{hartshorne2013algebraic}.
At a point $Q\in\operatorname{Spec}(\mathcal R/P)$, the stalk of the sheaf of regular functions is denoted by $\mathcal O_Q(\mathcal R/P)$; it identifies regular functions defined near $Q$ whenever they agree on a smaller neighborhood, and is given by
\begin{equation}
\mathcal O_Q(\mathcal R/P)=(\mathcal R/P)_Q.
\end{equation}
Here $(\mathcal R/P)_Q$ denotes the local ring at the prime ideal $Q$, obtained by making all elements outside $Q$ invertible~\cite{hartshorne2013algebraic}.
Since $\mathcal R/P$ is an integral domain, $\operatorname{Spec}(\mathcal R/P)$ is an integral scheme.
Its sheaf of rational functions is therefore a constant sheaf, and all nonempty open sets share the same function field
\begin{equation}
K(\mathcal R/P)=\operatorname{Frac}(\mathcal R/P),
\end{equation}
where $\operatorname{Frac}(\mathcal R/P)$ denotes the field of fractions of $\mathcal R/P$~\cite{hartshorne2013algebraic}.

When the first two conditions of the criterion hold, the Krull dimension of $\mathcal R/P$ is $1$, so $\operatorname{Spec}(\mathcal R/P)$ is a one-dimensional affine curve~\cite{atiyah2018introduction,hartshorne2013algebraic}.
To discuss the zeros and poles of rational functions on this curve, we first construct its projective closure and then normalize that closure.

Take a finitely generated, standard graded $\mathbb F_2$-algebra $\mathcal C_P$ that is an integral domain, and write
\begin{equation}
\mathcal C_P=\bigoplus_{d\ge0}S_d,\
S_0=\mathbb F_2,\
S_+=\bigoplus_{d>0}S_d.
\end{equation}
Here $S_d$ is the homogeneous component of degree $d$, including the zero element, and ``standard graded'' means that $\mathcal C_P$ is generated by elements of degree $1$.
We denote the projective scheme associated with the graded ring $\mathcal C_P$ by $\operatorname{Proj}(\mathcal C_P)$.
Its points correspond to the homogeneous prime ideals of $\mathcal C_P$ that do not contain $S_+$~\cite{goertz2020algebraic}.
As in the affine case, we describe its topology and its sheaves of regular and rational functions.
When discussing points, subsets, or images, we use the same notation for the underlying topological spaces of these schemes.
For any homogeneous ideal $I\subseteq\mathcal C_P$, define the closed set
\begin{equation}
V_I(\mathcal C_P)=\{Q\in\operatorname{Proj}(\mathcal C_P):I\subseteq Q\}.
\end{equation}
These sets are the Zariski closed sets of the projective scheme.
For any homogeneous element $h\in\mathcal C_P$ of positive degree, define the basic open set
\begin{equation}
D_h(\mathcal C_P)=\{Q\in\operatorname{Proj}(\mathcal C_P):h\notin Q\}.
\end{equation}
These sets form a basis for its topology~\cite{goertz2020algebraic}.
Here $D_h(\mathcal C_P)$ is the standard projective open set usually denoted by $D_+(h)$.
The sheaves of regular and rational functions on the projective scheme are denoted by $\mathcal O(\mathcal C_P)$ and $\mathcal K(\mathcal C_P)$, respectively.
On a basic open set $D_h(\mathcal C_P)$, the ring of regular functions is
\begin{equation}
\mathcal O_h(\mathcal C_P)=(\mathcal C_P[h^{-1}])_0,
\end{equation}
the degree-zero part of the localization of $\mathcal C_P$ at $h$.
For a point $Q\in\operatorname{Proj}(\mathcal C_P)$, let $S_Q$ be the multiplicative set of all homogeneous elements outside $Q$. The local ring at this point is then
\begin{equation}
\mathcal O_Q(\mathcal C_P)=(S_Q^{-1}\mathcal C_P)_0.
\end{equation}
That is, we first make the homogeneous elements outside $Q$ invertible and then take the degree-zero part.
Its elements can be written as quotients $a/b$ of homogeneous elements of the same degree, with $b\notin Q$.
These descriptions of the standard open sets and local rings specify the sheaf of regular functions on the projective scheme~\cite{hartshorne2013algebraic}.

Since $\mathcal C_P$ is an integral domain, $\operatorname{Proj}(\mathcal C_P)$ is also an integral scheme~\cite{goertz2020algebraic}, and its sheaf of rational functions is likewise a constant sheaf.
All nonempty open sets share the function field
\begin{equation}
K(\mathcal C_P)=\operatorname{Frac}_0(\mathcal C_P).
\end{equation}
Here $\operatorname{Frac}_0(\mathcal C_P)$ denotes the field consisting of quotients of homogeneous elements of the same degree, with nonzero denominators~\cite{hartshorne2013algebraic}.

We require $\operatorname{Proj}(\mathcal C_P)$ to be a projective closure of $\operatorname{Spec}(\mathcal R/P)$.
In the model used here, this means that there is a homogeneous element $T\in\mathcal C_P$ of positive degree such that
\begin{equation}
(\mathcal C_P[T^{-1}])_0\cong\mathcal R/P,
\end{equation}
and $D_T(\mathcal C_P)$ is dense in $\operatorname{Proj}(\mathcal C_P)$.
This ring isomorphism identifies the open subscheme $D_T(\mathcal C_P)\subseteq\operatorname{Proj}(\mathcal C_P)$ with $\operatorname{Spec}(\mathcal R/P)$~\cite{goertz2020algebraic}.
Denote the corresponding open immersion by
\begin{equation}
\nu_{P,1}:\operatorname{Spec}(\mathcal R/P)\longrightarrow\operatorname{Proj}(\mathcal C_P).
\end{equation}
This open immersion identifies $\operatorname{Spec}(\mathcal R/P)$ with the dense open subscheme $D_T(\mathcal C_P)$.
Since this open set is dense, the function fields of the affine curve and its projective closure are also identified:
\begin{equation}
K(\mathcal R/P)\cong K(\mathcal C_P).
\end{equation}
Denote this identification by $\phi_{P,1}:K(\mathcal R/P)\longrightarrow K(\mathcal C_P)$.

There are several ways to construct the projective closure of an affine curve, and the choice does not affect whether the third condition of the criterion holds.
We describe a standard construction here. The Laurent polynomial ring $\mathcal R$ is isomorphic to the following quotient of an ordinary polynomial ring:
\begin{equation}
\mathcal R\cong\mathbb F_2[x,y,z,u]/\langle1+xyzu\rangle.
\end{equation}
Specifically, let $B=\mathbb F_2[x,y,z,u]$ and define the surjection $\rho:B\longrightarrow\mathcal R$ by sending $x,y,z$ to the elements with the same names and $u$ to $(xyz)^{-1}$.
Set $I=\rho^{-1}(P)$. Then $I$ is a prime ideal containing $\langle1+xyzu\rangle$, and
\begin{equation}
\mathcal R/P\cong B/I.
\end{equation}
Introduce a new variable $T$ and assign degree $1$ to each of $X,Y,Z,U,T$.
For any nonzero polynomial $h\in B$ of total degree $d$, define its homogenization with respect to $T$ by
\begin{equation}
\operatorname{hom}(h)=T^dh(X/T,Y/T,Z/T,U/T).
\end{equation}
Let $J$ be the homogenization of the entire ideal $I$~\cite{polstra2026zariski}, namely,
\begin{equation}
J=\langle\operatorname{hom}(h):h\in I\rangle\subseteq\mathbb F_2[X,Y,Z,U,T].
\end{equation}
Set
\begin{equation}
\mathcal C_P=\mathbb F_2[X,Y,Z,U,T]/J.
\end{equation}
Then $J$ is a homogeneous prime ideal, so $\mathcal C_P$ is an integral domain~\cite{polstra2026zariski}.
On $D_T(\mathcal C_P)$, the identifications $x=X/T$, $y=Y/T$, $z=Z/T$, and $u=U/T$ give
\begin{equation}
(\mathcal C_P[T^{-1}])_0\cong B/I\cong\mathcal R/P.
\end{equation}
Moreover, $D_T(\mathcal C_P)$ is dense in $\operatorname{Proj}(\mathcal C_P)$.
Thus, this $\mathcal C_P$ satisfies the requirements above and gives the desired projective closure~\cite{polstra2026zariski,goertz2020algebraic}.

The projective curve $\operatorname{Proj}(\mathcal C_P)$ need not be normal, so we also normalize it.
Let $\overline{\mathcal C}_P$ be the integral closure of $\mathcal C_P$ in its ordinary field of fractions:
\begin{equation}
\overline{\mathcal C}_P=\{h\in\operatorname{Frac}(\mathcal C_P):h\text{ is integral over }\mathcal C_P\}.
\end{equation}
Here $h$ is integral over $\mathcal C_P$ if it satisfies a monic polynomial equation with coefficients in $\mathcal C_P$.
The ring $\overline{\mathcal C}_P$ inherits a grading and is a finite $\mathcal C_P$-module~\cite{huneke2006integral,goertz2020algebraic}.
The corresponding scheme $\operatorname{Proj}(\overline{\mathcal C}_P)$ is the normalization of $\operatorname{Proj}(\mathcal C_P)$ and is a one-dimensional normal projective curve~\cite{huneke2006integral,goertz2020algebraic}.
Denote the normalization morphism by
\begin{equation}
\nu_{P,2}:\operatorname{Proj}(\overline{\mathcal C}_P)\longrightarrow\operatorname{Proj}(\mathcal C_P).
\end{equation}
This morphism is finite, birational, and surjective.
On points, it contracts homogeneous prime ideals of $\overline{\mathcal C}_P$ to $\mathcal C_P$ and defines a continuous closed map.
Over the normal open part of the original curve, $\nu_{P,2}$ is an isomorphism; a nonnormal point may have one or more preimages~\cite{goertz2020algebraic,hartshorne2013algebraic}.
Regular functions can be pulled back along this morphism, and the normalization may admit additional regular functions that are not regular on the original curve.
For any subset $E\subseteq\operatorname{Proj}(\mathcal C_P)$, we denote its inverse image under the map on points by $\nu_{P,2}^{-1}(E)$~\cite{goertz2020algebraic}.
Normalization does not change the function field of the curve~\cite{goertz2020algebraic}, so there is an isomorphism
\begin{equation}
\phi_{P,2}:K(\mathcal C_P)\longrightarrow K(\overline{\mathcal C}_P).
\end{equation}
Thus, rational functions on the affine curve, its projective closure, and its normalization can all be identified through a common function field.
For any closed point $Q\in\operatorname{Proj}(\overline{\mathcal C}_P)$, the local ring $\mathcal O_Q(\overline{\mathcal C}_P)$ embeds naturally into the function field $K(\overline{\mathcal C}_P)$, and its field of fractions is precisely this function field.
Since the curve is normal and one-dimensional, the local ring at this closed point is a discrete valuation ring~\cite{hartshorne2013algebraic,atiyah2018introduction}.
Choose a local parameter $t$, that is, a generator of the maximal ideal of this local ring.
Every nonzero rational function $h\in K(\overline{\mathcal C}_P)$ has a unique decomposition
\begin{equation}
h=vt^n,\ v\in\mathcal O_Q(\overline{\mathcal C}_P)^\times,\ n\in\mathbb Z.
\end{equation}
The integer $n$ is independent of the choice of local parameter.
We therefore define the order of $h$ at $Q$ by
\begin{equation}
\operatorname{ord}_Q(h)=n.
\end{equation}
If $n>0$, then $h$ has a zero of order $n$ at $Q$; if $n<0$, then $h$ has a pole of order $-n$ at $Q$; and if $n=0$, then $h$ is invertible in the local ring at that point.
This definition gives a discrete valuation with values in $\mathbb Z$~\cite{atiyah2018introduction}.
For any nonzero rational functions $h_1,h_2$, these orders satisfy
\begin{equation}
\begin{aligned}
&\operatorname{ord}_Q(h_1h_2)=\operatorname{ord}_Q(h_1)+\operatorname{ord}_Q(h_2),\\
&\operatorname{ord}_Q(h_1/h_2)=\operatorname{ord}_Q(h_1)-\operatorname{ord}_Q(h_2),\\
&\operatorname{ord}_Q(h_1+h_2)\ge\min\{\operatorname{ord}_Q(h_1),\operatorname{ord}_Q(h_2)\}.
\end{aligned}
\end{equation}

We can now discuss rational functions on $\operatorname{Proj}(\overline{\mathcal C}_P)$.
Consider $x,y,z\in\mathcal R$, and denote the quotient map and the natural embedding into the field of fractions by
\begin{equation}
\eta_P:\mathcal R\longrightarrow\mathcal R/P,\ \tau_P:\mathcal R/P\longrightarrow K(\mathcal R/P).
\end{equation}
Let $\bar x,\bar y,\bar z$ denote their images under these maps and the function-field isomorphisms:
\begin{equation}
\begin{aligned}
\bar x&=(\phi_{P,2}\circ\phi_{P,1}\circ\tau_P\circ\eta_P)(x),\\
\bar y&=(\phi_{P,2}\circ\phi_{P,1}\circ\tau_P\circ\eta_P)(y),\\
\bar z&=(\phi_{P,2}\circ\phi_{P,1}\circ\tau_P\circ\eta_P)(z).
\end{aligned}
\end{equation}
Since $x,y,z$ are invertible in $\mathcal R$, these images are nonzero rational functions on the normal projective curve.
On the original affine scheme, $x,y,z$ are invertible regular functions, and their inverses are also regular.
Consequently, after pullback to the normalization, $\bar x,\bar y,\bar z$ remain invertible on the entire preimage of the affine part and have no zeros or poles there~\cite{atiyah2018introduction,hartshorne2013algebraic}.
Define
\begin{equation}
S_P=\operatorname{Proj}(\overline{\mathcal C}_P)\setminus\nu_{P,2}^{-1}\!\left(\nu_{P,1}\!\left(\operatorname{Spec}(\mathcal R/P)\right)\right).
\end{equation}
Thus, $S_P$ is the full preimage, under normalization, of the boundary of the projective closure relative to the original affine curve.
The functions $\bar x,\bar y,\bar z$ can therefore have zeros or poles only in $S_P$~\cite{atiyah2018introduction,hartshorne2013algebraic}.
The set $S_P$ consists of finitely many closed points~\cite{hartshorne2013algebraic}.
Writing $S_P=\{Q_1,\ldots,Q_{|S_P|}\}$, define
\begin{equation}
V_P=
\begin{pmatrix}
\operatorname{ord}_{Q_1}(\bar x)&\operatorname{ord}_{Q_1}(\bar y)&\operatorname{ord}_{Q_1}(\bar z)\\
\vdots&\vdots&\vdots\\
\operatorname{ord}_{Q_{|S_P|}}(\bar x)&\operatorname{ord}_{Q_{|S_P|}}(\bar y)&\operatorname{ord}_{Q_{|S_P|}}(\bar z)
\end{pmatrix}.
\end{equation}
Each row of the matrix records the orders of the three coordinate functions at one boundary point.
In the criterion, $\operatorname{rank}V_P$ denotes the rank of this integer matrix over $\mathbb Q$, which is also equal to its rank over $\mathbb R$.

For practical calculations of the order of a rational function at a closed point, the properties of a discrete valuation ring allow us to express it in terms of the length of a quotient module over the local ring~\cite{atiyah2018introduction}.
Write $\mathcal O_Q=\mathcal O_Q(\overline{\mathcal C}_P)$.
For any $h\in K(\overline{\mathcal C}_P)^\times$, we have
\begin{equation}
\operatorname{ord}_Q(h)=
\begin{cases}
\operatorname{length}_{\mathcal O_Q}
(\mathcal O_Q/h\mathcal O_Q),
&h\in\mathcal O_Q,\\[2mm]
-\operatorname{length}_{\mathcal O_Q}
(\mathcal O_Q/h^{-1}\mathcal O_Q),
&h\notin\mathcal O_Q.
\end{cases}
\end{equation}
The second case corresponds to a pole, for which $h^{-1}\in\mathcal O_Q$.
When $h$ is a unit in the local ring, $\mathcal O_Q/h\mathcal O_Q=0$, and both the module length and the order are $0$.
Here $\operatorname{length}_{\mathcal O_Q}(M)$ denotes the length of $M$ as an $\mathcal O_Q$-module.
If there is a composition series
\begin{equation}
0=M_0\subsetneq M_1\subsetneq\cdots\subsetneq M_n=M,
\end{equation}
such that each quotient $M_i/M_{i-1}$ is a simple module, meaning that it is nonzero and has no nontrivial submodules, then $\operatorname{length}_{\mathcal O_Q}(M)=n$~\cite{atiyah2018introduction}. The Jordan--H\"older theorem ensures that any two composition series have the same length, so the module length is well defined~\cite{atiyah2018introduction}.


\section{Details of proofs}
\label{sec:sufficiency_necessity}

In this section, we establish Eq.~\eqref{eq:criterion} as a necessary and sufficient condition for a uniform logarithmic lower bound on the logical energy barrier of the periodic TB code family.
Recent work shows that a logarithmic logical energy barrier yields partial self-correction under local Pauli Davies dynamics through an appropriate temperature-dependent choice of system size~\cite{gu2026layer}.
Our criterion therefore provides a computable means of certifying PSCQSs among TB code families.

We begin with the preliminaries in Subsec.~\ref{sub:pre} and two supporting theorems in Subsec.~\ref{sub:Supp1} needed for the proof in Subsecs.~\ref{sec:sufficiency} 
and~\ref{sec:necessity}, followed by proofs of sufficiency and necessity in Subsecs.~\ref{sec:sufficiency} and~\ref{sec:necessity}, 
respectively. 
In Subsec.~\ref{sub:psc_result}, we prove that every TB code family constitutes a PSCQS if it satisfies Eq.~\eqref{eq:criterion}.
Subsection~\ref{sec:proofs_of_lemmas} collects the supporting lemmas, together with their proofs.

We consider the Hamiltonian in Eq.~(\ref{eq:stabilizer_hamiltonian}), generated by two Laurent polynomials $f,g\in\mathcal R$. For a system of linear size $L$, the number of qubits is $n_L=2L^3$. A code constitutes a PSCQS if there exists a recovery procedure such that, for any fixed error tolerance, one can choose a temperature-dependent finite system size for which the memory time exhibits super-Arrhenius growth, with no restriction on the computational complexity of recovery. Throughout the proof, we consider the widely used thermal dynamics reviewed in Sec.~\ref{sec:psc_codes}.

\subsection{Preliminaries}
\label{sub:pre}

In this subsection, we define the energy barrier and
show how the energy-barrier condition reduces to topological-order conditions and uniform no-string conditions. 

\subsubsection{1. Energy barrier}

For clarity, we review the definition of the energy barrier~\cite{bravyi2009no,bravyi2013quantum}.
Fix a periodic size $L$ and disregard overall phases of Pauli operators throughout.
Let $|\psi_0\rangle$ be any ground state.
After a Pauli operator $E$ is applied, the state $E|\psi_0\rangle$ remains an eigenstate of the Hamiltonian, with the eigenvalues of exactly those stabilizer generators in $\operatorname{syn}(E)$ flipped from $+1$ to $-1$.
The excitation energy created by $E$ is
\begin{equation}
\mathcal E(E)=2J|\operatorname{syn}(E)|.
\label{eq:pauli_excitation_energy}
\end{equation}
We consider sequences of single-qubit Pauli operations starting from the identity.

A finite sequence $\gamma=(E_0,E_1,\ldots,E_m)$ is called a path implementing $E$ if
\begin{equation}
E_0=I,\ E_m=E,\ E_{j+1}=P_jE_j,
\end{equation}
where each $P_j$ is a Pauli operator acting on a single physical qubit.
The energy barrier of a path $\gamma$ is the maximum excitation energy encountered along it,
\begin{equation}
\Delta(\gamma)=\max_{0\leq j\leq m}\mathcal E(E_j)=2J\max_{0\leq j\leq m}|\operatorname{syn}(E_j)|.
\label{eq:path_energy_barrier}
\end{equation}
The energy barrier of a Pauli operator $E$ is defined as the minimum energy barrier over all paths implementing it~\cite{bravyi2013quantum}.
Thus, $\Delta(E)$ is the minimum peak excitation energy that must be reached to implement $E$ starting from a ground state.
For a finite code with $k_L>0$, the energy barrier of the code is the minimum over all nontrivial logical Pauli operators,
\begin{equation}
\Delta_L=\min_{E\in\mathcal C(\mathcal S_L)\setminus\mathcal S_L}
\Delta(E).
\label{eq:code_energy_barrier}
\end{equation}

$\mathcal C(\mathcal S_L)$ denotes the Pauli centralizer of the stabilizer group.
For the polynomial calculations, write $\mathcal R_L=\mathcal R/\langle x^L+1,y^L+1,z^L+1\rangle$ and let $\pi_L:\mathcal R\to\mathcal R_L$ be the quotient map.
For $h\in\mathcal R$, $h^T$ denotes the inversion $h(x^{-1},y^{-1},z^{-1})$.
We represent the translated $X$- and $Z$-type check generators by $(\pi_L(f)m\mid\pi_L(g)m)$ and $(\pi_L(g^T)m\mid\pi_L(f^T)m)$, respectively, where $m$ labels their cell.
With this convention, the excitation energy can also be computed directly using the polynomial representation.
Decompose an arbitrary Pauli operator as $E_XE_Z$, and use the polynomial representations $(\varepsilon_1^X\mid\varepsilon_2^X)$ and $(\varepsilon_1^Z\mid\varepsilon_2^Z)$.
The support of $h$, denoted by $\operatorname{supp}(h)$, is the set of monomials with nonzero coefficients.
The size of the support of a syndrome polynomial equals the number of stabilizer defects of the corresponding type.
\begin{equation}
\mathcal E(E)=2J\left|\operatorname{supp}\!\left(\pi_L(f^T)\varepsilon_1^Z+\pi_L(g^T)\varepsilon_2^Z\right)\right|+
2J\left|\operatorname{supp}\!\left(\pi_L(g)\varepsilon_1^X+\pi_L(f)\varepsilon_2^X\right)\right|.
\label{eq:polynomial_excitation_energy}
\end{equation}

\subsubsection{2. Topological order conditions and uniform no-strings condition}
\label{subsub:condition}

We now introduce two topological-order conditions, TQO1 and TQO2, and the uniform no-strings 
condition (UNSC)~\cite{bravyi2011energy}. Together, these conditions imply a lower bound on 
the energy barrier, as established in Ref.~\cite{bravyi2011energy}. For clarity, we restate this result in the following lemma.
\begin{lemma}
	\label{lem:tqo_unsc_barrier}
	If a code family satisfies TQO1, TQO2, and UNSC, and there is a constant $\eta>0$, 
	independent of $L$, such that $L_{\mathrm{tqo}}\ge L^\eta$, then its energy barrier 
	satisfies $\Delta_L=\Omega(\log n_L)$.
\end{lemma}
Here, $L_{\mathrm{tqo}}$ is a topological scale, and we choose 
$L_{\mathrm{tqo}}=L_{\mathrm{tqo}}(L)<L/4$. 
Before stating these conditions, we introduce the necessary notation.

We equip the periodic lattice $\mathbb Z_L^3$ with the $\ell_\infty$ metric.
For any point $a$ and $r>0$, let $N_r(a)$ denote the cubic box centered at $a$, consisting of all points at distance at most $r$ from $a$.
The distance between two cubic boxes is defined as the distance between their centers.
For any set of points $A$, let $N_r(A)$ denote its $r$-neighborhood, consisting of all points whose minimum distance to $A$ is at most $r$.
For a nonempty local defect cluster $S$, let $C(S)$ be a smallest cubic box containing it.

With this notation, we restate the three conditions from Ref.~\cite{bravyi2011energy} as follows.

\begin{definition}[TQO1]
\label{def:family_tqo1}
A code family satisfies TQO1 if, for all sufficiently large sizes, every Pauli operator $E$ supported in a cubic box of side length $L_{\mathrm{tqo}}$ satisfies
\begin{equation}
\operatorname{syn}(E)=\varnothing\ \Longrightarrow\ E\in\mathcal S_L.
\end{equation}
\end{definition}

A defect cluster $S$ is called neutral if there exists a Pauli operator $E$, supported in some cubic box of side length $L_{\mathrm{tqo}}$, that creates the cluster in isolation, i.e., $\operatorname{syn}(E)=S$.
Otherwise, $S$ is called charged.
The empty cluster is created by the identity operator and is therefore neutral.

\begin{definition}[TQO2]
\label{def:family_tqo2}
A code family satisfies TQO2 if there is a constant $r_0>0$, independent of $L$, such that, for all sufficiently large sizes, every nonempty neutral cluster $S$ can be created in isolation by a Pauli operator supported in the $r_0$-neighborhood of its bounding box.
\end{definition}

\begin{definition}[UNSC]
\label{def:family_unsc}
A code family satisfies UNSC if there is a constant $\alpha\ge1$, independent of $L$, such that the following holds for all sufficiently large sizes.
Consider any $E$ supported in a cubic box of side length $L_{\mathrm{tqo}}$, whose full syndrome is contained in two disjoint cubic boxes $A_1,A_2$, each of side length $\rho\ge1$.
If the distance between $A_1$ and $A_2$ exceeds $\alpha\rho$, then the defect clusters contained in $A_1,A_2$ are both neutral.
\end{definition}

We now reduce TQO1 and TQO2 for the finite TB code family to the topological-order condition (TOC) for the infinite TB code.
Let $\mathcal S_\infty$ denote the group of finite products of these local stabilizer generators.

\begin{definition}[TOC]
\label{def:infinite_toc}
The infinite TB code satisfies TOC if every finitely supported Pauli operator $E$ satisfies
\begin{equation}
\operatorname{syn}_\infty(E)=\varnothing\ \Longrightarrow\ E\in\mathcal S_\infty,
\end{equation}
where $\operatorname{syn}_\infty$ denotes the syndrome with respect to all checks on the infinite lattice.
\end{definition}

This TOC is the exactness condition in Haah's terminology.
His results on local topological order and the local creation of neutral clusters can be stated in our notation as following~\cite{haah2013commuting}.
\begin{lemma}
\label{lem:toc_family_tqo}
If the infinite TB code defined by fixed $f,g$ satisfies TOC, then there is a constant $c_{\mathrm{tqo}}\in(0,1/4)$, independent of $L$, such that the corresponding TB code family satisfies TQO1 and TQO2 at the topological scale $L_{\mathrm{tqo}}=c_{\mathrm{tqo}}L$.
\end{lemma}

Our previous work established a coprimality criterion for TOC~\cite{xu2026topological}: for the TB codes considered here, the infinite code satisfies TOC if and only if $\gcd(f,g)=1$.
Thus, the coprimality condition in the criterion already ensures TOC for the infinite TB code. 
Lemma~\ref{lem:toc_family_tqo} then gives TQO1 and TQO2 for the finite code family at a scale $L_{\mathrm{tqo}}=\Theta(L)$.
Together with Lemma~\ref{lem:tqo_unsc_barrier}, this reduces our task to establishing UNSC for the TB code family at such a topological scale.

We next reduce UNSC for the code family to UNSC for the infinite TB code.
On the infinite lattice, we use the $\ell_\infty$ metric and the convention that distances between cubic boxes are measured between their centers.
A finite defect cluster $S$ is neutral if there exists a finitely supported Pauli operator $E$ such that $\operatorname{syn}_\infty(E)=S$; otherwise, it is charged.

\begin{definition}[UNSC for an infinite code]
\label{def:infinite_unsc}
An infinite TB code satisfies UNSC if there is a constant $\alpha_\infty\ge1$ such that the following holds.
For any finitely supported Pauli operator $E$ whose full syndrome is contained in two disjoint cubic boxes $A_1,A_2$, each of side length $\rho\ge1$, if the distance between their centers exceeds $\alpha_\infty\rho$, then the defect clusters in the two boxes are both neutral.
\end{definition}

To turn neutral clusters in the infinite code into neutral clusters in the finite code, we need to control the supports of the operators that create them.
The local generation result gives the following lemma~\cite{haah2013commuting}.
\begin{lemma}
\label{lem:infinite_local_creation}
If the infinite TB code defined by fixed $f,g$ satisfies TOC, then there is a constant $r_0>0$, depending only on $f,g$, such that every nonempty neutral cluster $S$ can be created in isolation by a Pauli operator $F_S$ supported in $N_{r_0}(C(S))$, i.e.,
\begin{equation}
\operatorname{syn}_\infty(F_S)=S,\ \operatorname{supp}(F_S)\subseteq N_{r_0}(C(S)).
\end{equation}
\end{lemma}

\begin{lemma}
\label{lem:unsc_infinite_to_family}
If the infinite TB code satisfies TOC and UNSC, then the corresponding TB code family also satisfies UNSC at the topological scale $L_{\mathrm{tqo}}=c_{\mathrm{tqo}}L$ given by Lemma~\ref{lem:toc_family_tqo}.
\end{lemma}

\begin{proof}
Suppose, for a contradiction, that the code family fails to satisfy UNSC.
Then there is a sequence of sizes $L$ tending to infinity and Pauli operators $E_L$, each supported in a cubic box $B_L$ of side length $L_{\mathrm{tqo}}$, whose full syndromes consist of two defect clusters $S_{1,L},S_{2,L}$.
The clusters lie in disjoint cubic boxes $A_{1,L},A_{2,L}$ of side length $\rho_L>0$, at least one cluster is charged, and the distance $D_L$ between the box centers satisfies
\begin{equation}
\frac{D_L}{\rho_L}\longrightarrow\infty.
\end{equation}
Both clusters are nonempty: otherwise, $E_L$ itself would create the other cluster in isolation within the topological scale, making both clusters neutral.
Since the periodic $\ell_\infty$ distance satisfies $D_L\le L$, we have
\begin{equation}
\frac{\rho_L}{L}\le\frac{1}{(D_L/\rho_L)}
\longrightarrow0.
\end{equation}

Choose a fixed constant $w$ such that every check is supported within the $w$-neighborhood of the cell indexing it.
Then $\operatorname{syn}(E_L)\subseteq N_w(B_L)$.
Because both endpoint clusters are nonempty, both endpoint boxes are contained in $N_{\rho_L+w}(B_L)$.
Thus, $E_L$, the two endpoint boxes, and the full supports of all checks intersecting $E_L$ lie in $N_{\rho_L+2w}(B_L)$.
Using $\rho_L=o(L)$ and $c_{\mathrm{tqo}}<1/4$, for all sufficiently large $L$ the side length of this box satisfies
\begin{equation}
L_{\mathrm{tqo}}+2\rho_L+4w<L/2.
\end{equation}
The box therefore admits an isometric lift to the infinite lattice, yielding $\widetilde{E}_L$, $\widetilde A_{i,L}$, and $\widetilde S_{i,L}$ with the same endpoint-box side lengths and center distance, and
\begin{equation}
\operatorname{syn}_\infty(\widetilde{E}_L)=\widetilde S_{1,L}\sqcup\widetilde S_{2,L}.
\end{equation}

For sufficiently large $L$, we have $D_L/\rho_L>\alpha_\infty$.
UNSC for the infinite code implies that both lifted defect clusters are neutral.
By Lemma~\ref{lem:infinite_local_creation}, each $\widetilde S_{i,L}$ can be created in isolation by a Pauli operator $\widetilde{F}_{i,L}$ supported in the $r_0$-neighborhood of $C(\widetilde S_{i,L})$.
Since each cluster lies in a box of side length $\rho_L$, each creation operator is supported in a box of side length at most $\rho_L+2r_0$.

Because $\rho_L=o(L)$, for sufficiently large $L$ we have
\begin{equation}
\rho_L+2r_0<L_{\mathrm{tqo}}.
\end{equation}
Moreover, each creation operator and the full supports of the checks intersecting it fit inside a box of side length at most $\rho_L+2r_0+4w<L/2$.
Projecting these objects back to the finite lattice with period $L$ gives Pauli operators $F_{i,L}$ whose supports remain inside cubic boxes of side length $L_{\mathrm{tqo}}$ and satisfy
\begin{equation}
\operatorname{syn}(F_{i,L})=S_{i,L},\qquad i=1,2.
\end{equation}
Both endpoint clusters in the original finite code are therefore neutral, contradicting the assumption that at least one is charged.
Hence, the code family satisfies UNSC.
\end{proof}

\subsection{Two supporting theorems}
\label{sub:Supp1}

In this subsection, we express UNSC for an infinite code as a polynomial condition in the quotient ring.

\begin{theorem}[Polynomial formulation of UNSC]
	\label{thm:unsc_polynomial}
	An infinite TB code satisfies UNSC if and only if there is a constant $\alpha\ge1$, depending only on $f,g$, such that, for every positive integer $\rho$, every pair $s_1,s_2\in\mathcal R$ satisfying $\operatorname{Deg}(s_i)\le\rho$, and every monomial $r\in M$,
	\begin{equation}
		\pi(s_1)=\pi(r)\pi(s_2)\ne0\quad\Longrightarrow\frac{\operatorname{Deg}(r)}{\rho}\le\alpha.
		\label{eq:unsc_polynomial}
	\end{equation}
\end{theorem}

\begin{proof}
Represent the sites of the infinite lattice by the monomial set $M=\{x^ay^bz^c\mid(a,b,c)\in\mathbb Z^3\}$, with $x^ay^bz^c$ corresponding to the site $(a,b,c)$.
For a nonzero Laurent polynomial $h$, define
\begin{equation}
	\operatorname{Deg}(h)=\max_{x^ay^bz^c\in\operatorname{supp}(h)}\max\{|a|,|b|,|c|\},
	\label{eq:laurent_Deg}
\end{equation}
and set $\operatorname{Deg}(0)=0$.
The $\ell_\infty$ distance between two sites $m_1,m_2\in M$ is $\operatorname{Deg}(m_1^{-1}m_2)$.
For a positive integer $\rho$ and $m\in M$, define the cubic set of lattice sites centered at $m$, with side length $2\rho+1$, by
\begin{equation}
	\mathcal B_\rho(m)=\{mx^ay^bz^c\mid a,b,c\in\mathbb Z,\ |a|,|b|,|c|\le\rho\}.
\end{equation}
The monomial $m$ labels the center of the box.
The distance between two such boxes is
\begin{equation}
	\operatorname{dist}\bigl(\mathcal B_\rho(m_1),\mathcal B_\rho(m_2)\bigr)=\operatorname{Deg}(m_1^{-1}m_2).
	\label{eq:cube_center_distance}
\end{equation}

A TB code is a CSS code, so the two syndrome types can be treated separately.
We first consider defects of Z-type checkers, which are created by X-type errors $E_X$ expressed
as $(\varepsilon_1^X \mid \varepsilon_2^X)$ with syndrome polynomial $g\varepsilon_1^X+f\varepsilon_2^X$.
The two sectors have an $X,Z$ duality: for any $a,b\in\mathcal R$,
\begin{equation}
	(f^Ta+g^Tb)^T=fa^T+gb^T.
\end{equation}
The right-hand side is the syndrome polynomial of the X-type error $(b^T\mid a^T)$.
Thus, spatial inversion and exchange of the two qubit blocks relate the two sectors while preserving finite support, box side lengths, center distances, and cluster neutrality.
It therefore suffices to prove UNSC for the sector of Z-type checker defects.
For a general Pauli error, apply this result to the two CSS sectors separately. 
Multiplying the two finitely supported creation operators at each endpoint then creates its full defect cluster in isolation.

Let $\pi:\mathcal R\to\mathcal R/\langle f,g\rangle$ be the quotient map, and write $\pi(h)$ for the residue class of $h$.
A finite cluster of Z-type defects represented by a polynomial $s$ is neutral if and only if there exist $\varepsilon_1,\varepsilon_2\in\mathcal R$ such that $s=g\varepsilon_1+f\varepsilon_2$.
Thus,
\begin{equation}
	s\text{ is neutral}\Longleftrightarrow s\in \langle f,g\rangle\Longleftrightarrow \pi(s)=0,
	\label{eq:charge_quotient}
\end{equation}
whereas $\pi(s)\ne0$ represents a charged cluster.

Now consider any finitely supported $E_X$ whose full syndrome lies in two disjoint boxes $\mathcal B_\rho(r_1),\mathcal B_\rho(r_2)$.
Translate each endpoint cluster to the reference box $\mathcal B_\rho(1)$ and denote the resulting polynomials by $s_1,s_2$. 
Then
\begin{equation}
	g\varepsilon_1+f\varepsilon_2=r_1s_1+r_2s_2.
\end{equation}
Set $r=r_1^{-1}r_2$, multiply the equation by $r_1^{-1}$, and pass to the quotient to obtain
\begin{equation}
	\pi(s_1)=\pi(r)\pi(s_2).
	\label{eq:two_cluster_quotient}
\end{equation}
Since $\pi(r)$ is invertible, the two endpoints are either both neutral or both charged.
The latter case is exactly $\pi(s_1)=\pi(r)\pi(s_2)\ne0$, and $\operatorname{Deg}(r)$ is the distance between the box centers.
\end{proof}

Next, we provide the following theorem, which plays an essential role in the proof of sufficiency of
the algebraic criterion.

\begin{theorem}[Boundary valuation bound for charged clusters]
	\label{thm:branch_valuation_bound}
	Let $\mathcal R=\mathbb F_2[x^{\pm1},y^{\pm1},z^{\pm1}]$ and suppose that $f,g\in\mathcal R$ satisfy $\langle f,g\rangle\ne\mathcal R$ and $\gcd(f,g)=1$.
	There is a constant $C\ge0$, depending only on $f,g$, such that, for every positive integer $\rho$, every monomial $r\in M$, and each $s_1,s_2\in\mathcal R$ satisfying
	\begin{equation}
		\pi(s_1)=\pi(r)\pi(s_2)\ne0,\ \operatorname{Deg}(s_1),\operatorname{Deg}(s_2)\le\rho,
		\label{eq:unsc_endpoint_assumption}
	\end{equation}
	one can find at least one $P\in\operatorname{Min}(\langle f,g\rangle)$ such that
	\begin{equation}
		\max_{Q\in S_P}\left|\operatorname{ord}_Q(\bar r)\right|\le C\rho.
		\label{eq:boundary_order_bound}
	\end{equation}
\end{theorem}

\begin{proof} 
	Since the proof requires numerous definitions, we divide it into two parts. 
	The first part introduces the necessary definitions, and the second uses them to establish the propositions.
	The proof requires many lemmas, which are proved in Subsec.~\ref{sec:proofs_of_lemmas}.
	
	\begin{center}
	\textit{Part 1}
	\end{center}

	For each $P\in\operatorname{Min}(\langle f,g\rangle)$, define the localized ring and its natural map by
	\begin{equation}
		\mathcal A_P=\left(\mathcal R/\langle f,g\rangle\right)_{P/\langle f,g\rangle},\ \chi_P:\mathcal R/\langle f,g\rangle\longrightarrow\mathcal A_P.
		\label{eq:localization_map_chi}
	\end{equation}
	Its unique maximal ideal is
	\begin{equation}
		\mathfrak m_P=\left(P/\langle f,g\rangle\right)\mathcal A_P.
		\label{eq:local_maximal_ideal}
	\end{equation}
	Throughout this first part, fix $P\in\operatorname{Min}(\langle f,g\rangle)$ and use the natural quotient and fraction-field maps $\eta_P$ and $\tau_P$ from Sec.~\ref{sec:criterion}, together with the local ring $\mathcal A_P$, localization map $\chi_P$, and maximal ideal $\mathfrak m_P$ in Eqs.~\eqref{eq:localization_map_chi} and \eqref{eq:local_maximal_ideal}.
	Write the corresponding residue-field quotient map as
	\begin{equation}
		\mu_P:\mathcal A_P\longrightarrow\mathcal A_P/\mathfrak m_P.
		\label{eq:local_residue_quotient}
	\end{equation}
	Lemma~\ref{lem:generic_residue_isomorphism} gives a natural ring isomorphism
	\begin{equation}
		\theta_P:\mathcal A_P/\mathfrak m_P\rightarrow\text{Frac}(\mathcal R/P),
		\label{eq:generic_residue_field}
	\end{equation}
	such that, for $a,b\in\mathcal R$ with $b\notin P$,
	\begin{equation}
		(\theta_P\circ\mu_P)
		\left(\frac{\pi(a)}{\pi(b)}\right)=
		\frac{(\tau_P\circ\eta_P)(a)}{(\tau_P\circ\eta_P)(b)}.
		\label{eq:local_fraction_residue}
	\end{equation}
	Lemma~\ref{lem:residue_maps_commute} further shows that these maps form the commutative diagram
	\begin{equation}
		\begin{tikzcd}[ampersand replacement=\&, column sep=1.4em, row sep=2em]
			\mathcal R \arrow[r,"\eta_P"] \arrow[d,"\pi"']
			\& \mathcal R/P \arrow[r,"\tau_P"]
			\& \text{Frac}(\mathcal R/P) \\
			\mathcal R/\langle f,g\rangle \arrow[r,"\chi_P"']
			\& \mathcal A_P \arrow[r,"\mu_P"']
			\& \mathcal A_P/\mathfrak m_P \arrow[u,"\theta_P"']
		\end{tikzcd}
		\label{eq:residue_commuting_diagram}
	\end{equation}
	
	We next consider the quotient layers of the localized ring.
	For $j\ge0$, set $\mathfrak m_P^0=\mathcal A_P$ and denote the natural quotient map of modules by
	\begin{equation}
		\kappa_{P,j}:\mathfrak m_P^j\longrightarrow\mathfrak m_P^j/\mathfrak m_P^{j+1}.
		\label{eq:layer_quotient_map}
	\end{equation}
	Denote the dimension of $\mathfrak m_P^j/\mathfrak m_P^{j+1}$ over $\text{Frac}(\mathcal R/P)$ by $D_{P,j}$ and fix a basis for each nonzero layer,
	\begin{equation}
		e_1^{P,j},\ldots,e_{D_{P,j}}^{P,j}.
	\end{equation}
	The associated coordinate isomorphism and its $i$-th component are defined by
	\begin{equation}
		\lambda^{P,j}:\mathfrak m_P^j/\mathfrak m_P^{j+1}\rightarrow\text{Frac}^{D_{P,j}}(\mathcal R/P),\quad
		\lambda_i^{P,j}:\mathfrak m_P^j/\mathfrak m_P^{j+1}\rightarrow\text{Frac}(\mathcal R/P),
		\label{eq:layer_coordinate_isomorphism}
	\end{equation}
	where for any $h\in\mathfrak m_P^j/\mathfrak m_P^{j+1}$, which can be written as
	$h=\sum_{i=1}^{D_{P,j}} c_i e_i^{P,j}$, we have
	\begin{equation}
		\lambda^{P,j}(h)= \begin{pmatrix}
			c_1 \\
			c_2 \\
			\vdots \\
			c_{D_{P,j}}
		\end{pmatrix}
	\end{equation}
	and $\lambda^{P,j}_i(h)=c_i$.
	Each $\lambda_i^{P,j}$ is $\text{Frac}(\mathcal R/P)$-linear.
	
	Define an ideal of $\mathcal R$ by
	\begin{equation}
		J_{P,j}=\left\{h\in\mathcal R\mid(\chi_P\circ\pi)(h)\in\mathfrak m_P^j\right\}.
		\label{eq:layer_preimage_ideal}
	\end{equation}
	Define the composite coordinate map by
	\begin{equation}
		U_{P,j,i}=\lambda_i^{P,j}\circ\kappa_{P,j}\circ\chi_P\circ\pi\mid_{J_{P,j}}:J_{P,j}\longrightarrow\text{Frac}(\mathcal R/P).
		\label{eq:coordinate_composite_domain}
	\end{equation}
	Since $\mathcal R$ is Noetherian, $J_{P,j}$ is finitely generated. Fix a set of generators
	\begin{equation}
		J_{P,j}=\langle h_1^{P,j},\ldots,h_{\omega_{P,j}}^{P,j}\rangle.
	\end{equation}
	For each generator, choose $a_{P,j,i,k},b_{P,j,i,k}\in\mathcal R/P$ with $b_{P,j,i,k}\ne0$ such that
	\begin{equation}
		(\lambda_i^{P,j}\circ\kappa_{P,j}\circ\chi_P\circ\pi)(h_k^{P,j})=
		\frac{\tau_P(a_{P,j,i,k})}{\tau_P(b_{P,j,i,k})}.
		\label{eq:generator_coordinate_fraction}
	\end{equation}
	Define the common denominator
	\begin{equation}
		\Lambda_{P,j,i}=\prod_{k=1}^{\omega_{P,j}}\tau_P(b_{P,j,i,k}).
		\label{eq:coordinate_common_denominator}
	\end{equation}
	Define the map $\xi_{P,j,i}:J_{P,j}\longrightarrow\mathcal R/P$ given by
	\begin{equation}
		\xi_{P,j,i}(h)=\tau_P^{-1}\left(\Lambda_{P,j,i}(\lambda_i^{P,j}\circ\kappa_{P,j}\circ\chi_P\circ\pi)(h)\right).
		\label{eq:cleared_coordinate_map}
	\end{equation}
	Here $\tau_P^{-1}$ denotes only the inverse of the injection $\tau_P$ on its image $\tau_P(\mathcal R/P)$; Lemma~\ref{lem:cleared_coordinate_linear} ensures that this definition is well defined.
	
	All these bases, generators, and denominators depend only on the fixed indices $P,j,i$, independently of the endpoints to which the maps are subsequently applied.
	
		\begin{center}
		\textit{Part 2}
	\end{center}
	
	By Lemma~\ref{lem:minimal_localization_detection}, there exists a $P\in\operatorname{Min}(\langle f,g\rangle)$ such that
	\begin{equation}
		(\chi_P\circ\pi)(s_1)\ne0,\ (\chi_P\circ\pi)(s_2)\ne0.
		\label{eq:nonzero_localized_endpoints}
	\end{equation}
	Lemma~\ref{lem:common_nonzero_layer} provides an integer $N_P\ge1$ with $\mathfrak m_P^{N_P}=0$ and an index $0\le j<N_P$ such that
	\begin{equation}
		(\chi_P\circ\pi)(s_1),\ (\chi_P\circ\pi)(s_2)\in\mathfrak m_P^j\setminus\mathfrak m_P^{j+1}.
		\label{eq:common_nonzero_layer}
	\end{equation}
	Thus $s_1,s_2\in J_{P,j}$ and their images in this quotient layer are nonzero.
	
	Since $\pi(s_1)=\pi(rs_2)$, Lemma~\ref{lem:coordinate_composite_linear} gives, for every coordinate $i$,
	\begin{equation}
		U_{P,j,i}(s_1)=U_{P,j,i}(rs_2)=(\tau_P\circ\eta_P)(r)U_{P,j,i}(s_2).
		\label{eq:coordinate_endpoint_relation}
	\end{equation}
	Choose $i$ such that $U_{P,j,i}(s_2)\ne0$.
	As $r$ is a unit, $U_{P,j,i}(s_1)$ is also nonzero.
	Using the fixed common denominator in Eq.~\eqref{eq:cleared_coordinate_map}, we obtain
	\begin{equation}
		(\tau_P\circ\eta_P)(r)=\frac{U_{P,j,i}(s_1)}{U_{P,j,i}(s_2)}=\frac{(\tau_P\circ\xi_{P,j,i})(s_1)}{(\tau_P\circ\xi_{P,j,i})(s_2)}.
		\label{eq:layer_coordinate_fraction}
	\end{equation}
	By Lemma~\ref{lem:filtered_representative_degree}, there exist $A,B\in\mathcal R$ such that
	\begin{equation}
		\eta_P(A)=\xi_{P,j,i}(s_1),\ \eta_P(B)=\xi_{P,j,i}(s_2),\ \operatorname{Deg}(A),\operatorname{Deg}(B)\le\rho+C_{P,j,i}.
		\label{eq:polynomial_A_B}
	\end{equation}
	Both quotient images are nonzero, and substitution yields
	\begin{equation}
		(\tau_P\circ\eta_P)(r)=\frac{(\tau_P\circ\eta_P)(A)}{(\tau_P\circ\eta_P)(B)}.
		\label{eq:polynomial_fraction_r}
	\end{equation}
	Applying the function-field isomorphisms $\phi_{P,1}$ and $\phi_{P,2}$ gives $\bar r=\bar A/\bar B$.
	Lemma~\ref{lem:curve_order_rules} therefore implies
	\begin{equation}
		\max_{Q\in S_P}\left|\operatorname{ord}_Q(\bar r)\right|\le C'_{P,j,i}\rho.
	\end{equation}
	For each fixed $P$, only the finitely many layers $0\le j<N_P$ can be nonzero, and each has finite dimension $D_{P,j}$.
	Since there are only finitely many minimal primes, we can fix all bases and denominators in advance and set
	\begin{equation}
		C=\max_{\substack{P\in\operatorname{Min}(\langle f,g\rangle)\\0\le j<N_P\\1\le i\le D_{P,j}}}C'_{P,j,i},
		\label{eq:branch_uniform_valuation_constant}
	\end{equation}
	where only nonzero layers are included.
	This constant is finite, depends only on $f,g$, and is independent of $P,j,i,s_1,s_2,r,\rho$. Eq.~\eqref{eq:boundary_order_bound} is thus proved.
\end{proof}

\subsection{Sufficiency}
\label{sec:sufficiency}

In this subsection, we prove that if given $f,g\in\mathcal R$, 
$\langle f,g\rangle\ne\mathcal R$, $\gcd(f,g)=1$, and $\operatorname{rank}V_P=3$ for every $P\in\operatorname{Min}(\langle f,g\rangle)$,
then the TB code family $\{\mathcal S_L\}$ defined by $f,g$ admits a uniform logarithmic lower bound on the logical energy barrier.

The condition $\gcd(f,g)=1$ already ensures TOC for the infinite TB code~\cite{xu2026topological}. 
With the preceding reductions, it thus remains only to prove UNSC for the infinite code.

Based on Theorem~\ref{thm:branch_valuation_bound}, for every positive integer $\rho$, every monomial $r\in M$, 
and each $s_1,s_2\in\mathcal R$ satisfying
$
	\pi(s_1)=\pi(r)\pi(s_2)\ne0,\ \operatorname{Deg}(s_1),\operatorname{Deg}(s_2)\le\rho,
$
there exists a $P\in\operatorname{Min}(\langle f,g\rangle)$ such that
\begin{equation}
	\max_{Q\in S_P}\left|\operatorname{ord}_Q(\bar r)\right|\le C\rho.
\end{equation}
We now focus on this $P$ and write $r=x^ay^bz^c$ and $\bar r=(\phi_{P,2}\circ\phi_{P,1}\circ\tau_P\circ\eta_P)(r)$.
The additivity of valuations and the definition of $V_P$ give
\begin{equation}
V_P
\begin{pmatrix}
a\\
b\\
c
\end{pmatrix}=
\left(\operatorname{ord}_Q(\bar r)\right)_{Q\in S_P}.
\label{eq:valuation_matrix_translation}
\end{equation}
Since $\operatorname{rank}V_P=3$, there is a fixed left inverse $L_P$ satisfying $L_PV_P=I_3$.
Hence
\begin{equation}
\operatorname{Deg}(r)=\left\|(a,b,c)^T\right\|_\infty= \left\| L_P V_P (a,b,c)^T\right\|_\infty  \le  \|L_P\|_{\infty\to\infty}\max_{Q\in S_P}\left|\operatorname{ord}_Q(\bar r)\right|\le\|L_P\|_{\infty\to\infty}C\rho.
\label{eq:translation_degree_bound}
\end{equation}
Note that for different $r\in M$ and $s_1,s_2\in\mathcal R$, the considered $P$ may be different.  
We thus define
\begin{equation}
	\alpha=C\max_{P\in\operatorname{Min}(\langle f,g\rangle)}\|L_P\|_{\infty\to\infty}.
	\label{eq:uniform_unsc_constant}
\end{equation}
As there are only finitely many minimal primes, so $\alpha$
is finite and depends only on $f,g$.
Thus every endpoint relation in Eq.~\eqref{eq:unsc_endpoint_assumption} obeys $\operatorname{Deg}(r)\le\alpha\rho$.
By Theorem~\ref{thm:unsc_polynomial}, the infinite TB code satisfies UNSC.

By Lemma~\ref{lem:unsc_infinite_to_family}, the finite TB code family also satisfies UNSC.
Together with TQO1 and TQO2 from Lemma~\ref{lem:toc_family_tqo}, Lemma~\ref{lem:tqo_unsc_barrier} gives $\Delta_L=\Omega(\log n_L)$.


\subsection{Necessity}
\label{sec:necessity}

In this subsection, we prove that if the TB code family $\{\mathcal S_L\}$ defined by $f,g\in\mathcal R$ admits a uniform logarithmic lower bound on the logical energy barrier, then $\langle f,g\rangle\ne\mathcal R$, $\gcd(f,g)=1$, and $\operatorname{rank}V_P=3$ for every $P\in\operatorname{Min}(\langle f,g\rangle)$.
We here prove the contrapositive, that is, if the algebraic criterion conditions are not satisfied, then
the corresponding code family has a bounded energy barrier.

We first consider a TB code on the infinite lattice $\mathbb Z^3$. 
If either TOC or the full-rank condition fails, we can construct a logical error channel whose energy barrier remains bounded independently of scale. 
The constructions use local operations of fixed support, so the same energy estimates apply to the corresponding error paths in finite codes.

If TOC fails, Definition~\ref{def:infinite_toc} guarantees the existence of a finite-support Pauli operator $E_0$ that commutes with all stabilizers but is not a product of finitely many local stabilizers.
Thus, $E_0$ is itself a local logical operator.
Producing it requires only a fixed number of single-qubit operations, and the number of defects along the entire path is bounded by a constant.
Local errors can therefore change the logical state, and increasing the system size cannot raise the energy barrier of this logical error path.

Now suppose that TOC holds, equivalently $\gcd(f,g)=1$, and that there exists at least one $P\in\operatorname{Min}(\langle f,g\rangle)$ such that $\operatorname{rank}V_P<3$.
The proper-ideal condition excludes the degenerate case in which none of the finite periodic codes encodes a logical qubit.
Using the convention for X-type errors, represent a finite-support Pauli operator by the polynomial pair $e=(\varepsilon_1\mid\varepsilon_2)\in\mathcal R^2$. 
Its syndrome polynomial is $g\varepsilon_1+f\varepsilon_2$.
We will construct a nontrivial translation $t$, a charged defect cluster $h$, and a local operator $e$ that transports $h$ to $th$.

To construct $t$, choose a nonzero integer vector $v\in\ker_{\mathbb Z}V_P$ and set $m=x^{v_x}y^{v_y}z^{v_z}$.
Here $\bar m$ denotes the image of $m$ on the normal projective curve introduced in Sec.~\ref{sec:criterion}.
Since $V_Pv=0$, $\bar m$ has no zeros or poles on the normal projective curve of the branch $P$, so $\bar m$ is a nonzero element of its finite constant field.
Let $q$ be the multiplicative order of this constant. 
Then
\begin{equation}
1+m^q\in P.
\label{eq:nec_binomial}
\end{equation}
Because $P$ is a minimal prime ideal, the image $(\chi_P\circ\pi)(1+m^q)$ in the local quotient ring $\mathcal A_P$ defined in Eq.~\eqref{eq:localization_map_chi} is nilpotent.
Choose a fixed integer $s\ge0$ such that $(\chi_P\circ\pi)((1+m^q)^{2^s})=0$, and define $t=m^{q2^s}$, so that
\begin{equation}
(\chi_P\circ\pi)(1+t)=0.
\label{eq:nec_translation}
\end{equation}
Here we use the characteristic-$2$ identity $(1+m^q)^{2^s}=1+m^{q2^s}$.

By the definition of localization, there exists $h\in\mathcal R$ such that $h\notin P$ and $h(1+t)\in\langle f,g\rangle$.
Thus there exist $\varepsilon_1,\varepsilon_2\in\mathcal R$ for which the finite-support operator $e=(\varepsilon_1\mid\varepsilon_2)$ satisfies
\begin{equation}
g\varepsilon_1+f\varepsilon_2=h(1+t).
\label{eq:nec_seed}
\end{equation}
Since $\langle f,g\rangle\subseteq P$, the condition $h\notin P$ implies $h\notin\langle f,g\rangle$. Thus, $h$ is a charged cluster that cannot be created alone by a finite-support operator.
The monomial $t$ is a unit, so every translate $t^jh$ also lies outside $\langle f,g\rangle$ and is therefore charged.
The operator $e$ creates the two clusters $h$ and $th$, thereby implementing one local transport step.

We now concatenate translated copies of this operator along $t$ to form
\begin{equation}
E_M=(1+t+\cdots+t^{M-1})(\varepsilon_1\mid\varepsilon_2),\ M\ge1.
\label{eq:nec_finite_segment}
\end{equation}
Its syndrome is
\begin{equation}
(1+t+\cdots+t^{M-1})h(1+t)=h(1+t^M).
\label{eq:nec_open_string}
\end{equation}
At each junction, the syndromes of neighboring segments cancel, leaving only the two charged endpoint clusters $h$ and $t^Mh$.
Counting each completed application of the fixed-support operator $e$ as one step, we have throughout the construction
\begin{equation}
|\operatorname{supp}(h(1+t^M))|\le 2|\operatorname{supp}(h)|.
\label{eq:nec_syndrome_bound}
\end{equation}
The string width is fixed by the support of $e$, whereas the separation between its charged endpoints grows with $M$. 
The charges can therefore be separated arbitrarily far while the number of defects remains bounded.

We now show that, for infinitely many finite periods, the segment in Eq.~\eqref{eq:nec_finite_segment} closes into a nontrivial logical operator with an energy barrier bounded independently of system size.
In the preceding construction, the integer kernel vector $v$ can be chosen primitive by dividing its components by their greatest common divisor; this preserves $V_Pv=0$.
We use this choice and write $\ell=q2^s$, so that $t=m^\ell$.

We first choose a period that preserves the charge of $h$.
By the Hilbert Nullstellensatz~\cite{atiyah2018introduction}, the conditions $f,g\in P$ and $h\notin P$, with $P$ prime, ensure that, in some finite extension $\mathbb F_{2^{r_*}}$ of $\mathbb F_2$, there exist nonzero coordinates $\alpha_x,\alpha_y,\alpha_z$ such that
\begin{equation}
f(\alpha_x,\alpha_y,\alpha_z)=g(\alpha_x,\alpha_y,\alpha_z)=0,\quad h(\alpha_x,\alpha_y,\alpha_z)\ne0.
\label{eq:nec_finite_field_point}
\end{equation}
Set $L_0=\ell(2^{r_*}-1)$.
For every $L=nL_0$, $n\ge1$, all three coordinates have $L$th power equal to $1$, so evaluation at this point respects the periodic relations.
We claim that
\begin{equation}
\pi_L(h)\notin\langle\pi_L(f),\pi_L(g)\rangle
\quad(L=nL_0).
\label{eq:nec_finite_charge}
\end{equation}
Otherwise, there would exist $a,b\in\mathcal R$ such that
\begin{equation}
h+af+bg\in\langle1+x^L,1+y^L,1+z^L\rangle.
\end{equation}
Evaluating this relation at $(\alpha_x,\alpha_y,\alpha_z)$ would give $h(\alpha_x,\alpha_y,\alpha_z)=0$, a contradiction.
Thus no periodic Pauli operator, regardless of its support, can create this cluster in isolation.
In particular, it is charged according to the finite-size definition using $L_{\mathrm{tqo}}$.
Every translate $t^jh$ remains charged because $\pi_L(t)$ is a unit.

For $L=nL_0$, take $M=\frac{L}{\ell}=n(2^{r_*}-1)$.
Then $t^M=m^L=x^{Lv_x}y^{Lv_y}z^{Lv_z}$ and $\pi_L(t^M)=1$.
Consequently, the finite X-type operator $E_{L,M}\longleftrightarrow\pi_L(E_M)$ has
\begin{equation}
\operatorname{syn}_Z(E_{L,M})\longleftrightarrow\pi_L\!\left(h(1+t^M)\right)=0.
\label{eq:nec_periodic_closure}
\end{equation}
The total displacement is $L v$, corresponding to one complete winding along the primitive direction $m$.
The two charged endpoint clusters therefore meet and cancel.

To prove that this closed operator is a nontrivial logical operator, suppose instead that it is a stabilizer.
Since it is X-type, there would exist $\lambda\in\mathcal R$ with
\begin{equation}
\pi_L(E_M)=\bigl(\pi_L(f\lambda)\mid\pi_L(g\lambda)\bigr).
\end{equation}
Because $v$ is primitive, it can be completed to an integer basis of $\mathbb Z^3$.
Let $b_2,b_3$ be the monomials corresponding to the other two basis vectors.
This change of lattice basis gives
\begin{equation}
\mathcal R=\mathbb F_2[m^{\pm1},b_2^{\pm1},b_3^{\pm1}],\quad\langle 1+x^L, 1+y^L, 1+z^L\rangle=\langle1+m^L,1+b_2^L,1+b_3^L\rangle.
\end{equation}
Hence the assumed finite stabilizer relation lifts to the exact identity in $\mathcal R^2$
\begin{equation}
E_M=(f\lambda\mid g\lambda)+(1+m^L)(\eta_1\mid\eta_2)+(1+b_2^L)(\xi_1\mid\xi_2)+(1+b_3^L)(\zeta_1\mid\zeta_2),
\label{eq:nec_stabilizer_lift}
\end{equation}
for some $\eta_1,\eta_2,\xi_1,\xi_2,\zeta_1,\zeta_2\in\mathcal R$.
Applying the syndrome map to Eq.~\eqref{eq:nec_stabilizer_lift}, and using the exact identity $g(E_M)_1+f(E_M)_2=h(1+m^L)$ in $\mathcal R$, yields
\begin{equation}
(1+m^L)h=(1+m^L)(g\eta_1+f\eta_2)+(1+b_2^L)(g\xi_1+f\xi_2)+(1+b_3^L)(g\zeta_1+f\zeta_2).
\label{eq:nec_syndrome_lift}
\end{equation}
Now impose only the two transverse periodic relations $1+b_2^L=1+b_3^L=0$, leaving $m$ as a free Laurent variable.
In this intermediate quotient, $1+m^L$ is a non-zero-divisor: multiplying any nonzero finite Laurent expansion by it leaves a nonzero coefficient at the highest resulting power of $m$.
We can therefore cancel $1+m^L$ to obtain
\begin{equation}
h\equiv g\eta_1+f\eta_2
\pmod{\langle1+b_2^L,1+b_3^L\rangle}.
\end{equation}
It follows that
\begin{equation}
\begin{aligned}
h&\in\langle f,g,1+b_2^L,1+b_3^L\rangle\subseteq\langle f,g\rangle+\langle 1+x^L, 1+y^L, 1+z^L\rangle,
\end{aligned}
\end{equation}
contradicting Eq.~\eqref{eq:nec_finite_charge}.
Thus $E_{L,M}$ has zero syndrome and is not a stabilizer, so it is a nontrivial logical operator.
Rank deficiency therefore produces constant-barrier logical errors for infinitely many finite periods and excludes a uniform logarithmic energy-barrier lower bound over all sufficiently large encoded periods.

\subsection{Connection to partial self-correction}
\label{sub:psc_result}

We have established the necessary and sufficient criterion for a uniform logarithmic logical energy-barrier lower bound.
We now connect this result to partial self-correction.

To establish that a quantum LDPC code family constitutes a PSCQS, it suffices to show that its logical operators have a logarithmically growing energy barrier.
This result was proved in Refs.~\cite{williamson2026partial,gu2026layer} and is restated in the following lemma.

\begin{lemma}
	\label{lem:barrier_pscqs}
	Let $\{\mathcal C_n\}$ be a qLDPC code family with energy barrier $\Delta_n$.
	If $\Delta_n=\Omega\left(\log n\right)$, then the family constitutes a PSCQS.
\end{lemma}

Because $f,g$ are fixed, the check weights, the number of checks involving each qubit, and the interaction range are all uniformly bounded independently of $L$. 
Hence, $\{\mathcal S_L\}$ is a qLDPC code family.
As every TB code family satisfying Eq.~\eqref{eq:criterion} 
exhibits logarithmic logical energy-barrier lower bound, it constitutes a PSCQS.
The RG decoder constructed in Sec.~\ref{sec:efficient_decoders} corrects all Pauli errors below a logarithmic energy-barrier threshold in polynomial time, therefore yielding PSCQMs.

We have shown that, when the criterion is not satisfied, 
there exists an energy barrier upper bound for infinitely many system sizes. This does not rule out the possibility that other system sizes define a subfamily whose logical energy barriers satisfy a logarithmic lower bound. If such a subfamily exists, the criterion is not necessary for PSCQM, since its definition permits a temperature-dependent choice of system size $L^*$. 
We illustrate this possibility with a model containing a rank-deficient component.

Define
\begin{equation}
f=f_0,\quad g=hg_0,\quad f_0=1+x+y,\quad g_0=1+xy+z,\quad h=1+z+z^2.
\end{equation}
The ideal $\langle f,g\rangle$ has two minimal prime components, $\langle f_0,g_0\rangle$ and $\langle f_0,h\rangle$. 
The boundary matrix of the latter has rank $2$, so the criterion is not satisfied. 
Nevertheless, a suitable choice of the period $L$ establishes a correspondence between the $(f,g)$ and $(f_0,g_0)$ models. 
As shown in the main text, the latter satisfies the sufficient criterion.
Let
\begin{equation}
\mathcal R_L=\mathcal R/\langle 1+x^L,1+y^L,1+z^L\rangle.
\end{equation}
When $3\nmid L$, we have $\gcd(h,1+z^L)=1$, so $h$ is invertible in $\mathcal R_L$. 
For an arbitrary $X$ error
\begin{equation}
E=(\epsilon_1\mid\epsilon_2)
\end{equation}
in the $(f,g)$ model, define its corresponding error in the $(f_0,g_0)$ model by
\begin{equation}
T_X(E)=(h\epsilon_1\mid\epsilon_2).
\end{equation}
The two errors produce exactly the same syndrome:
\begin{equation}
\operatorname{syn}_{Z,0}(T_X(E))=g_0h\epsilon_1+f_0\epsilon_2=g\epsilon_1+f\epsilon_2=\operatorname{syn}_{Z}(E).
\end{equation}
This correspondence also allows us to compare error growth paths. 
Since $h$ contains only three fixed local monomials, each single-qubit $X$ operation in the $(f,g)$ model corresponds to $X$ operations on at most three neighboring qubits in the $(f_0,g_0)$ model. 
Applying these operations one at a time maps any single-qubit error path in the $(f,g)$ model to a single-qubit error path in the $(f_0,g_0)$ model. At the end of each operation block corresponding to an original step, the two syndromes are identical. 
Within each block, the number of additional defects is bounded by a constant $C$, independent of $L$ and the path length. 
Thus, the peak defect number along the mapped path increases by at most $C$. 
The argument for $Z$ errors is analogous.

Furthermore, to verify that nontrivial logical errors are preserved, note that any $X$ stabilizer satisfies
\begin{equation}
T_X\bigl(u(f\mid g)\bigr)=hu(f_0\mid g_0).
\end{equation}
Since $hu$ ranges over all of $\mathcal R_L$ as $u$ does, $T_X$ maps the stabilizer spaces of the two models bijectively onto each other. 
Together with the syndrome identity, this implies that $E$ is a nontrivial logical error if and only if $E_0=T_X(E)$ is a nontrivial logical error. 
The same property holds in the $Z$ sector.

Therefore, any path implementing a logical error in the $(f,g)$ model yields a path implementing a logical error in the $(f_0,g_0)$ model, with the peak defect number increased by at most a constant $C$. 
It follows that
\begin{equation}
\Delta_L\ge\Delta_{0,L}-C.
\end{equation}
Hence, on the subfamily of periods satisfying $3\nmid L$ and $k_L>0$, the $(f,g)$ model also has an energy barrier of $\Omega(\log L)$.
Finally, an efficient decoder for errors with low energy barriers can also be obtained through the above correspondence.

\subsection{Supporting lemmas}
\label{sec:proofs_of_lemmas}

In this subsection, we present the lemmas used in the preceding proof, together with their proofs.

\begin{lemma}
\label{lem:generic_residue_isomorphism}
There is a unique ring isomorphism
\begin{equation}
\theta_P:\mathcal A_P/\mathfrak m_P\rightarrow\text{Frac}(\mathcal R/P)
\end{equation}
such that, for all $a,b\in\mathcal R$ with $b\notin P$,
\begin{equation}
(\theta_P\circ\mu_P)\left(\frac{(\chi_P\circ\pi)(a)}{(\chi_P\circ\pi)(b)}\right)=
\frac{(\tau_P\circ\eta_P)(a)}{(\tau_P\circ\eta_P)(b)}.
\label{eq:appendix_local_fraction_formula}
\end{equation}
\end{lemma}

\begin{proof}
The map $\tau_P\circ\eta_P:\mathcal R\longrightarrow \text{Frac}(\mathcal R/P)$ is a ring homomorphism.
Since $\ker\eta_P=P$ and $\tau_P$ is injective, we have
$\ker(\tau_P\circ\eta_P)=P$. 
Hence $\langle f,g\rangle\subseteq\ker(\tau_P\circ\eta_P)$.
By the universal property of quotient rings, there is a unique ring homomorphism~\cite{atiyah2018introduction}
\begin{equation}
\varphi_P:\mathcal R/\langle f,g\rangle\longrightarrow \text{Frac}(\mathcal R/P)
\end{equation}
satisfying
\begin{equation}
\varphi_P\circ\pi=\tau_P\circ\eta_P.
\end{equation}

Let $s$ be any element inverted in the localization.
It has the form $s=\pi(b)$ with $b\notin P$.
Thus
\begin{equation}
\varphi_P(s)=(\tau_P\circ\eta_P)(b)\neq0.
\end{equation}
Since $\text{Frac}(\mathcal R/P)$ is a field, $\varphi_P(s)$ is a unit.
The universal property of localization therefore gives a unique ring homomorphism~\cite{kemper2011course}
\begin{equation}
\Phi_P:\mathcal A_P\longrightarrow \text{Frac}(\mathcal R/P)
\end{equation}
satisfying
\begin{equation}
\Phi_P\circ\chi_P=\varphi_P.
\end{equation}
Consequently, for all $a,b\in\mathcal R$ with $b\notin P$,
\begin{equation}
\Phi_P\left(\frac{(\chi_P\circ\pi)(a)}{(\chi_P\circ\pi)(b)}\right)=
\frac{(\tau_P\circ\eta_P)(a)}{(\tau_P\circ\eta_P)(b)}.
\label{eq:local_to_function_field_map}
\end{equation}

Every $q\in\text{Frac}(\mathcal R/P)$ can be written as $q=\tau_P(\alpha)/\tau_P(\beta)$, where $\alpha,\beta\in\mathcal R/P$ and $\beta\neq0$.
Since $\eta_P$ is surjective, we may choose $a,b\in\mathcal R$
with $\eta_P(a)=\alpha$ and $\eta_P(b)=\beta$.
Then $b\notin P$, and
Eq.~\eqref{eq:local_to_function_field_map} gives
\begin{equation}
q=\Phi_P\left(\frac{(\chi_P\circ\pi)(a)}{(\chi_P\circ\pi)(b)}\right).
\end{equation}
Thus $\Phi_P$ is surjective.

On the other hand, for $b\notin P$,
\begin{equation}
\Phi_P\left(\frac{(\chi_P\circ\pi)(a)}{(\chi_P\circ\pi)(b)}\right)=0\Longleftrightarrow a\in P.
\end{equation}
Therefore
\begin{equation}
\ker\Phi_P=\mathfrak m_P.
\end{equation}

By the first isomorphism theorem for rings, there is a unique ring isomorphism~\cite{eisenbud1995commutative}
\begin{equation}
\theta_P:\mathcal A_P/\mathfrak m_P\rightarrow\text{Frac}(\mathcal R/P)
\end{equation}
satisfying
\begin{equation}
\theta_P\circ\mu_P=\Phi_P.
\end{equation}
Together with Eq.~\eqref{eq:local_to_function_field_map}, this gives
Eq.~\eqref{eq:appendix_local_fraction_formula}.
\end{proof}

The isomorphism constructed above is compatible with the two natural quotient maps from $\mathcal R$.
We now verify the compatibility needed for the commutative diagram.

\begin{lemma}
\label{lem:residue_maps_commute}
The maps $\pi,\chi_P,\mu_P,\theta_P,\eta_P,\tau_P$ satisfy
\begin{equation}
\theta_P\circ\mu_P\circ\chi_P\circ\pi=
\tau_P\circ\eta_P.
\end{equation}
\end{lemma}

\begin{proof}
Let $h\in\mathcal R$.
By the definitions of $\chi_P$ and $\Phi_P=\theta_P\circ\mu_P$,
\begin{equation}
\begin{aligned}
(\theta_P\circ\mu_P\circ\chi_P\circ\pi)(h)=\Phi_P\left(\frac{\chi_P\circ\pi(h)}{\chi_P\circ\pi(1)}\right)=(\tau_P\circ\eta_P)(h).
\end{aligned}
\end{equation}
Since this identity holds for every $h\in\mathcal R$, the two composite maps coincide.
\end{proof}

The quotient layer $\mathfrak m_P^j/\mathfrak m_P^{j+1}$ is a vector space over the residue field $\mathcal A_P/\mathfrak m_P$~\cite{atiyah2018introduction}.
Through the isomorphism $\theta_P$, it is also a vector space over $\text{Frac}(\mathcal R/P)$.
Since $\mathcal A_P$ is Noetherian, the ideal $\mathfrak m_P^j$ is finitely generated, and the images of its generators span this quotient layer.
The layer is therefore a finite-dimensional $\text{Frac}(\mathcal R/P)$-vector space.
As the inverse image of the ideal $\mathfrak m_P^j$ under the ring homomorphism $\chi_P\circ\pi$, $J_{P,j}$ is an ideal of $\mathcal R$.

We next establish the compatibility of the layer coordinate maps with multiplication in the original ring.
Lemma~\ref{lem:coordinate_composite_linear} shows that, for every $t\in\mathcal R$ and $h\in J_{P,j}$,
\begin{equation}
(\lambda_i^{P,j}\circ\kappa_{P,j}\circ\chi_P\circ\pi)(th)=(\tau_P\circ\eta_P)(t)(\lambda_i^{P,j}\circ\kappa_{P,j}\circ\chi_P\circ\pi)(h).
\label{eq:coordinate_composite_linearity}
\end{equation}
This is the key step that turns the translation monomial $r$ into a scalar in the function field.

\begin{lemma}[Module compatibility of the composite layer coordinate map]
\label{lem:coordinate_composite_linear}
Fix $P,j$. With the scalar action and the coordinates in a fixed basis defined below,
for every $1\le i\le D_{P,j}$,
$t\in\mathcal R$, and $h\in J_{P,j}$, we have
\begin{equation}
(\lambda_i^{P,j}\circ\kappa_{P,j}\circ\chi_P\circ\pi)(th)=(\tau_P\circ\eta_P)(t)(\lambda_i^{P,j}\circ\kappa_{P,j}\circ\chi_P\circ\pi)(h).
\label{eq:appendix_coordinate_composite_linearity}
\end{equation}
\end{lemma}

\begin{proof}
As $\mathfrak m_P$ is a maximal ideal, $\mathcal A_P/\mathfrak m_P$ is a field.
By Lemma~\ref{lem:generic_residue_isomorphism},
$\theta_P:\mathcal A_P/\mathfrak m_P\rightarrow\operatorname{Frac}(\mathcal R/P)$ is a field isomorphism.

First define the scalar action of $\mathcal A_P/\mathfrak m_P$ on $\mathfrak m_P^j/\mathfrak m_P^{j+1}$.
For $\alpha\in \mathcal A_P/\mathfrak m_P$ and $w\in \mathfrak m_P^j/\mathfrak m_P^{j+1}$, choose representatives $a\in\mathcal A_P$ and $v\in\mathfrak m_P^j$ with $\mu_P(a)=\alpha$ and $\kappa_{P,j}(v)=w$, and set $\alpha\cdot w:=\kappa_{P,j}(av)$.
Since $\mathfrak m_P^j$ is an ideal, $av\in\mathfrak m_P^j$, so the right-hand side is defined.
If $a',v'$ are another pair of representatives of the same elements, then $a-a'\in\mathfrak m_P$ and $v-v'\in\mathfrak m_P^{j+1}$.
Hence
\begin{equation}
av-a'v'=(a-a')v+a'(v-v')\in\mathfrak m_P^{j+1},
\end{equation}
and therefore $\kappa_{P,j}(av)=\kappa_{P,j}(a'v')$.
Thus $\alpha\cdot w:=\kappa_{P,j}(av)$ is independent of both representatives.
Associativity, distributivity, and the identity axiom follow from the ring operations in $\mathcal A_P$, making $\mathfrak m_P^j/\mathfrak m_P^{j+1}$ a vector space over $\mathcal A_P/\mathfrak m_P$.

Next define the scalar action of $\operatorname{Frac}(\mathcal R/P)$ on the same layer: for $c\in \operatorname{Frac}(\mathcal R/P)$ and $w\in \mathfrak m_P^j/\mathfrak m_P^{j+1}$, set
\begin{equation}
c\star w:=\theta_P^{-1}(c)\cdot w.
\label{eq:coordinate_function_field_action}
\end{equation}
Since $\theta_P^{-1}$ preserves addition, multiplication, and the identity, this action makes $\mathfrak m_P^j/\mathfrak m_P^{j+1}$ a vector space over $\operatorname{Frac}(\mathcal R/P)$.
For every $a\in\mathcal A_P$ and $v\in\mathfrak m_P^j$, the two actions satisfy
\begin{equation}
(\theta_P\circ\mu_P)(a)\star\kappa_{P,j}(v)=\theta_P^{-1}\bigl(\theta_P(\mu_P(a))\bigr)\cdot\kappa_{P,j}(v)=\mu_P(a)\cdot\kappa_{P,j}(v)=\kappa_{P,j}(av).
\label{eq:coordinate_scalar_action_compatibility}
\end{equation}
Fix a basis of $\mathfrak m_P^j/\mathfrak m_P^{j+1}$ over $\operatorname{Frac}(\mathcal R/P)$ with respect to the $\star$ action, $e_1^{P,j},\ldots,e_{D_{P,j}}^{P,j}$.
For the unique expansion
\begin{equation}
w=\sum_{\ell=1}^{D_{P,j}}c_\ell\star e_\ell^{P,j},\ c_\ell\in \operatorname{Frac}(\mathcal R/P),
\end{equation}
define the coordinate map by
\begin{equation}
\lambda_i^{P,j}(w):=c_i.
\end{equation}
For every $c\in \operatorname{Frac}(\mathcal R/P)$, associativity and distributivity of the scalar action give
\begin{equation}
c\star w=\sum_{\ell=1}^{D_{P,j}}(cc_\ell)\star e_\ell^{P,j}.
\end{equation}
Uniqueness of the basis expansion then implies
\begin{equation}
\lambda_i^{P,j}(c\star w)=c\lambda_i^{P,j}(w).
\label{eq:coordinate_linearity_for_defined_action}
\end{equation}
The multiplication on the right-hand side is multiplication in $\operatorname{Frac}(\mathcal R/P)$.

Now let $t\in\mathcal R$ and $h\in J_{P,j}$, and write
\begin{equation}
a=(\chi_P\circ\pi)(t),\ v=(\chi_P\circ\pi)(h).
\end{equation}
By the definition of $J_{P,j}$, we have $v\in\mathfrak m_P^j$.
Since $\chi_P\circ\pi$ is a ring homomorphism and $\mathfrak m_P^j$ is an ideal,
\begin{equation}
(\chi_P\circ\pi)(th)=av\in\mathfrak m_P^j.
\end{equation}
Thus $th\in J_{P,j}$, so the composite map is defined on this input.
Lemma~\ref{lem:residue_maps_commute} also gives
\begin{equation}
(\theta_P\circ\mu_P)(a)=(\theta_P\circ\mu_P\circ\chi_P\circ\pi)(t)=(\tau_P\circ\eta_P)(t).
\end{equation}
Combining Eqs.~\eqref{eq:coordinate_scalar_action_compatibility} and~\eqref{eq:coordinate_linearity_for_defined_action}, we obtain
\begin{equation}
(\lambda_i^{P,j}\circ\kappa_{P,j}\circ\chi_P\circ\pi)(th)
=\lambda_i^{P,j}\bigl(\kappa_{P,j}(av)\bigr)
=(\tau_P\circ\eta_P)(t)(\lambda_i^{P,j}\circ\kappa_{P,j}\circ\chi_P\circ\pi)(h).
\end{equation}
This proves Eq.~\eqref{eq:appendix_coordinate_composite_linearity}.
\end{proof}

To pull the layer coordinates back from the function field to $\mathcal R/P$, we clear their denominators uniformly for each fixed coordinate.
In particular,
\begin{equation}
\Lambda_{P,j,i}(\lambda_i^{P,j}\circ\kappa_{P,j}\circ\chi_P\circ\pi)(J_{P,j})\subseteq\tau_P(\mathcal R/P),
\end{equation}
Lemma~\ref{lem:cleared_coordinate_linear} establishes that this map is well defined and is an $\mathcal R$-module homomorphism, with
\begin{equation}
\xi_{P,j,i}(th)=\eta_P(t)\xi_{P,j,i}(h).
\label{eq:cleared_coordinate_identity}
\end{equation}
The next lemma shows that the common denominator chosen above works for every element of $J_{P,j}$.

\begin{lemma}
\label{lem:cleared_coordinate_linear}
Fix $P,j,i$, together with the generators of $J_{P,j}$ and the common denominator $\Lambda_{P,j,i}$ chosen above. 
Then
\begin{equation}
\Lambda_{P,j,i}(\lambda_i^{P,j}\circ\kappa_{P,j}\circ\chi_P\circ\pi)(J_{P,j})\subseteq\tau_P(\mathcal R/P).
\label{eq:all_inputs_cleared}
\end{equation}
Consequently, Eq.~\eqref{eq:cleared_coordinate_map} defines a map $\xi_{P,j,i}:J_{P,j}\to\mathcal R/P$. This map is an $\mathcal R$-module homomorphism, namely,
\begin{equation}
\xi_{P,j,i}(h+h')=\xi_{P,j,i}(h)+\xi_{P,j,i}(h'),\quad
\xi_{P,j,i}(th)=\eta_P(t)\xi_{P,j,i}(h)
\label{eq:xi_module_linearity}
\end{equation}
for all $h,h'\in J_{P,j}$ and $t\in\mathcal R$.
\end{lemma}

\begin{proof}
Let $h\in J_{P,j}$. Since the fixed elements $h_1^{P,j}, \ldots,h_{\omega_{P,j}}^{P,j}$ generate $J_{P,j}$, there exist $t_k\in\mathcal R$ such that
\begin{equation}
h=\sum_{k=1}^{\omega_{P,j}}t_kh_k^{P,j}.
\end{equation}
This expansion need not be unique. 
By Lemma~\ref{lem:coordinate_composite_linear} and Eq.~\eqref{eq:generator_coordinate_fraction},
\begin{equation}
(\lambda_i^{P,j}\circ\kappa_{P,j}\circ\chi_P\circ\pi)(h)=\sum_{k=1}^{\omega_{P,j}}(\tau_P\circ\eta_P)(t_k)\frac{\tau_P(a_{P,j,i,k})}{\tau_P(b_{P,j,i,k})}.
\end{equation}
Multiplying by $\Lambda_{P,j,i}=\prod_k\tau_P(b_{P,j,i,k})$ gives
\begin{equation}
\Lambda_{P,j,i}(\lambda_i^{P,j}\circ\kappa_{P,j}\circ\chi_P\circ\pi)(h)=\tau_P\!\left(\sum_{k=1}^{\omega_{P,j}}\eta_P(t_k)a_{P,j,i,k}\prod_{\ell\ne k}b_{P,j,i,\ell}\right).
\label{eq:denominator_clearing_all_inputs}
\end{equation}
The element in parentheses belongs to $\mathcal R/P$, proving Eq.~\eqref{eq:all_inputs_cleared}. The left-hand side is determined by the map $\lambda_i^{P,j}\circ\kappa_{P,j}\circ\chi_P\circ\pi$ and is independent of the chosen generator expansion. Since $\tau_P$ is injective, it has a unique preimage in $\mathcal R/P$. Thus $\xi_{P,j,i}$ is well defined.

By the additivity of $\lambda_i^{P,j}\circ\kappa_{P,j}\circ\chi_P\circ\pi$,
\begin{equation}
(\tau_P\circ\xi_{P,j,i})(h+h')
=\Lambda_{P,j,i}(\lambda_i^{P,j}\circ\kappa_{P,j}\circ\chi_P\circ\pi)(h+h')
=(\tau_P\circ\xi_{P,j,i})(h)+(\tau_P\circ\xi_{P,j,i})(h').
\end{equation}
By Lemma~\ref{lem:coordinate_composite_linear},
\begin{equation}
(\tau_P\circ\xi_{P,j,i})(th)
=\Lambda_{P,j,i}(\lambda_i^{P,j}\circ\kappa_{P,j}\circ\chi_P\circ\pi)(th)
=(\tau_P\circ\eta_P)(t)(\tau_P\circ\xi_{P,j,i})(h).
\end{equation}
Since $\tau_P$ is an injective ring homomorphism, Eq.~\eqref{eq:xi_module_linearity} follows.
\end{proof}

Clearing denominators produces an element of $\mathcal R/P$.
Lemma~\ref{lem:filtered_representative_degree} further provides a constant $C_{P,j,i}\ge0$ such that, for every $h\in J_{P,j}$ with $\operatorname{Deg}(h)\le\rho$, there exists $t\in\mathcal R$ satisfying
\begin{equation}
\eta_P(t)=\xi_{P,j,i}(h),\ \operatorname{Deg}(t)\le\rho+C_{P,j,i}.
\label{eq:bounded_output_representative}
\end{equation}
The next lemma shows that, when the cube degree of the input Laurent polynomial is bounded, this quotient element admits a representative whose degree increases by at most a fixed constant.

\begin{lemma}
\label{lem:filtered_representative_degree}
For each fixed $P,j,i$, there is a nonnegative integer $C_{P,j,i}$ such that, for every $\rho\ge0$ and $s\in J_{P,j}$ with $\operatorname{Deg}(s)\le\rho$, there exists $t\in\mathcal R$ satisfying
\begin{equation}
\eta_P(t)=\xi_{P,j,i}(s),\
\operatorname{Deg}(t)\le\rho+C_{P,j,i}.
\label{eq:appendix_bounded_representative}
\end{equation}
\end{lemma}

\begin{proof}
Fix $P,j,i$.
For integers $d\ge0$, define
\begin{equation}
F_d\mathcal R=\{0\}\cup\{h\in\mathcal R\setminus\{0\}:\operatorname{Deg}(h)\le d\},\quad
F_dJ_{P,j}=J_{P,j}\cap F_d\mathcal R,\quad
F_d(\mathcal R/P)=\eta_P(F_d\mathcal R),
\label{eq:cube_degree_filtrations}
\end{equation}
and set the filtration pieces with negative indices to zero.
These filtrations are increasing and exhaustive.
Subadditivity of the cube degree gives
\begin{equation}
(F_a\mathcal R)(F_bJ_{P,j})\subseteq F_{a+b}J_{P,j}.
\end{equation}
Moreover, if $r\in F_a\mathcal R$ and $\eta_P(s)\in F_b(\mathcal R/P)$, then
\begin{equation}
\eta_P(rs)\in F_{a+b}(\mathcal R/P),
\end{equation}
so the target filtration is also compatible with the $\mathcal R$-module action.

The associated graded ring
\begin{equation}
\operatorname{gr}_F\mathcal R=\bigoplus_{d\ge0}F_d\mathcal R/F_{d-1}\mathcal R
\end{equation}
is generated by the leading symbols of the $26$ degree-$1$ monomials whose exponent vectors lie in
$\{-1,0,1\}^3\setminus\{(0,0,0)\}$.
Indeed, every Laurent monomial of degree $d$ can be expressed as a product of $d$ degree-$1$ monomials, with total filtration degree $d$.
Thus $\operatorname{gr}_F\mathcal R$ is a finitely generated $\mathbb F_2$-algebra and is Noetherian by the Hilbert basis theorem.

The degree filtration on $\mathcal R$ is therefore good.
Since
\begin{equation}
F_dJ_{P,j}=J_{P,j}\cap F_d\mathcal R,
\end{equation}
the induced filtration on $J_{P,j}$ is also good: $\operatorname{gr}_FJ_{P,j}$ is a graded submodule of $\operatorname{gr}_F\mathcal R$, which is Noetherian, so $\operatorname{gr}_FJ_{P,j}$ is finitely generated.

By Lemma~\ref{lem:cleared_coordinate_linear},
\begin{equation}
\xi_{P,j,i}:J_{P,j}\longrightarrow\mathcal R/P
\end{equation}
is a fixed $\mathcal R$-module homomorphism.
A fixed module homomorphism from a module with a good filtration to an exhaustively filtered module admits a finite shift in filtration degree.
Hence there is a nonnegative integer $C_{P,j,i}$ such that, for every $d\ge0$,
\begin{equation}
\xi_{P,j,i}(F_dJ_{P,j})
\subseteq
F_{d+C_{P,j,i}}(\mathcal R/P).
\label{eq:xi_filtration_shift}
\end{equation}

Now let $s\in J_{P,j}$ satisfy
$\operatorname{Deg}(s)\le\rho$.
Then $s\in F_\rho J_{P,j}$, and
Eq.~\eqref{eq:xi_filtration_shift} gives
\begin{equation}
\xi_{P,j,i}(s)
\in
F_{\rho+C_{P,j,i}}(\mathcal R/P)
=
\eta_P(F_{\rho+C_{P,j,i}}\mathcal R).
\end{equation}
Therefore there exists
$t\in F_{\rho+C_{P,j,i}}\mathcal R$
such that
\begin{equation}
\eta_P(t)=\xi_{P,j,i}(s).
\end{equation}
By the definition of $F_{\rho+C_{P,j,i}}\mathcal R$,
\begin{equation}
\operatorname{Deg}(t)
\le
\rho+C_{P,j,i}.
\end{equation}
as required.
\end{proof}

The preceding construction applies to any fixed branch. For a given pair of nonzero endpoints, we must also show that at least one minimal-prime branch preserves the nonvanishing of both endpoints.

\begin{lemma}
\label{lem:minimal_localization_detection}
Assume that $\langle f,g\rangle\ne\mathcal R$ and $\gcd(f,g)=1$.
If a monomial $r$ and elements $s_1,s_2\in\mathcal R$ satisfy
\begin{equation}
\pi(s_1)=\pi(r)\pi(s_2)\ne0,
\end{equation}
then there exists $P\in\operatorname{Min}(\langle f,g\rangle)$ such that
\begin{equation}
(\chi_P\circ\pi)(s_1)\ne0,\ (\chi_P\circ\pi)(s_2)\ne0.
\end{equation}
\end{lemma}

\begin{proof}
Write $\mathcal B=\mathcal R/\langle f,g\rangle$.
The Laurent polynomial ring $\mathcal R$ is a Noetherian unique factorization domain and is Cohen--Macaulay.
The assumptions $\langle f,g\rangle\ne\mathcal R$ and $\gcd(f,g)=1$ imply that $f,g$ are both nonzero nonunits.
Since $\mathcal R$ is a domain, $f$ is a non-zero-divisor.
If $gu\in(f)$, then $f\mid gu$; unique factorization and $\gcd(f,g)=1$ imply $f\mid u$.
Thus $g$ is also a non-zero-divisor in $\mathcal R/(f)$, so $f,g$ form a regular sequence.
The quotient of a Cohen--Macaulay ring by a regular sequence is again Cohen--Macaulay. 
Hence $\mathcal B$ is a Noetherian Cohen--Macaulay ring, and
\begin{equation}
\operatorname{Ass}(\mathcal B)=\operatorname{Min}(\mathcal B),
\end{equation}
so $\mathcal B$ has no embedded associated primes~\cite{eisenbud1995commutative}.

Let $Q(\mathcal B)$ be the total quotient ring obtained by localizing $\mathcal B$ at all its non-zero-divisors.
The natural map $\mathcal B\to Q(\mathcal B)$ is injective.
Indeed, if $u/1=0$, there is a non-zero-divisor $v$ with $vu=0$, which forces $u=0$.
For a Noetherian ring with no embedded associated primes, the total quotient ring admits the natural decomposition~\cite{cring2011project}
\begin{equation}
Q(\mathcal B)\cong\prod_{P\in\operatorname{Min}(\langle f,g\rangle)}\mathcal A_P.
\label{eq:total_quotient_decomposition}
\end{equation}
Here the product denotes the direct product of rings.
Consequently, there is an injection
\begin{equation}
\mathcal B
\hookrightarrow
\prod_{P\in\operatorname{Min}(\langle f,g\rangle)}\mathcal A_P,
\end{equation}
whose $P$th component is precisely $\chi_P$.

Since $\pi(s_1)\ne0$, its image in this product cannot vanish in every component.
Thus there exists $P$ such that $(\chi_P\circ\pi)(s_1)\ne0$.
On the same branch,
\begin{equation}
(\chi_P\circ\pi)(s_1)=(\chi_P\circ\pi)(r)(\chi_P\circ\pi)(s_2).
\end{equation}
The Laurent monomial $r$ is a unit in $\mathcal R$, so $(\chi_P\circ\pi)(r)$ is a unit in $\mathcal A_P$.
Since the left-hand side is nonzero, $(\chi_P\circ\pi)(s_2)\ne0$.
\end{proof}

Once such a branch is selected, its Artinian local structure places both endpoints in the same nonzero layer of the maximal-ideal filtration.

\begin{lemma}
\label{lem:common_nonzero_layer}
For every $P\in\operatorname{Min}(\langle f,g\rangle)$, there is an integer $N_P\ge1$, depending only on this branch, such that $\mathfrak m_P^{N_P}=0$.
If
\begin{equation}
(\chi_P\circ\pi)(s_1)\ne0,\quad
\pi(s_1)=\pi(r)\pi(s_2),
\end{equation}
and $r$ is a monomial, then there exists $0\le j<N_P$ such that
\begin{equation}
(\chi_P\circ\pi)(s_1),
(\chi_P\circ\pi)(s_2)
\in\mathfrak m_P^j\setminus\mathfrak m_P^{j+1}.
\end{equation}
\end{lemma}

\begin{proof}
The ring $\mathcal A_P$ is Noetherian. Its prime ideals correspond to the prime ideals of $\mathcal R/\langle f,g\rangle$ contained in $P/\langle f,g\rangle$. Since $P/\langle f,g\rangle$ is a minimal prime, $\mathcal A_P$ has only one prime ideal, namely $\mathfrak m_P$. 
Thus $\mathcal A_P$ is a zero-dimensional Noetherian local ring and hence an Artinian local ring. Its maximal ideal is nilpotent, so $\mathfrak m_P^{N_P}=0$ for some $N_P\ge1$~\cite{eisenbud1995commutative}.

The localized endpoint relation is
\begin{equation}
(\chi_P\circ\pi)(s_1)
=(\chi_P\circ\pi)(r)(\chi_P\circ\pi)(s_2).
\end{equation}
Since $r$ is a Laurent monomial, $(\chi_P\circ\pi)(r)$ is a unit. Both localized endpoints are therefore nonzero. 
Because $\mathfrak m_P^0=\mathcal A_P$ and $\mathfrak m_P^{N_P}=0$, there is a largest $j\in\{0,\ldots,N_P-1\}$ such that $(\chi_P\circ\pi)(s_2)\in\mathfrak m_P^j$. 
For every $q\ge0$, multiplication by the unit $(\chi_P\circ\pi)(r)$ and by its inverse preserves the ideal $\mathfrak m_P^q$, so
\begin{equation}
(\chi_P\circ\pi)(s_1)\in\mathfrak m_P^q
\ \Longleftrightarrow\
(\chi_P\circ\pi)(s_2)\in\mathfrak m_P^q.
\end{equation}
By maximality of $j$, both endpoints lie in $\mathfrak m_P^j\setminus\mathfrak m_P^{j+1}$.
\end{proof}

Finally, we prove the valuation estimate used in Theorem~\ref{thm:branch_valuation_bound}. This estimate converts a bound on the cube degree of Laurent representatives into bounds on their orders at the boundary points of the normal projective curve.

\begin{lemma}[Valuation bound for fractions of bounded degree]
\label{lem:curve_order_rules}
For each fixed $P,j,i$, there is a constant $C'_{P,j,i}\ge0$ such that, for every positive integer $\rho$ and all $A,B\in\mathcal R$ satisfying
\begin{equation}
\eta_P(A),\eta_P(B)\ne0,\ \operatorname{Deg}(A),\operatorname{Deg}(B)\le\rho+C_{P,j,i},
\end{equation}
we have, for every $Q\in S_P$,
\begin{equation}
\left|\operatorname{ord}_Q\left(\frac{\bar A}{\bar B}\right)\right|
\le C'_{P,j,i}\rho.
\label{eq:translation_valuation_bound}
\end{equation}
\end{lemma}

\begin{proof}
At every closed point $Q$ of the normal projective curve $\operatorname{Proj}(\bar{\mathcal C}_P)$, the local ring $\mathcal O_Q$ is a discrete valuation ring~\cite{eisenbud1995commutative}. 
The corresponding valuation therefore satisfies
\begin{equation}
\begin{aligned}
&\operatorname{ord}_Q(uv)=\operatorname{ord}_Q(u)+\operatorname{ord}_Q(v),\\
&\operatorname{ord}_Q(u/v)=\operatorname{ord}_Q(u)-\operatorname{ord}_Q(v),\\
&\operatorname{ord}_Q(u+v)\ge\min\{\operatorname{ord}_Q(u),\operatorname{ord}_Q(v)\},
\end{aligned}
\label{eq:discrete_valuation_rules}
\end{equation}
where the last inequality uses the convention $\operatorname{ord}_Q(0)=+\infty$.

For $Q\in S_P$, define
\begin{equation}
\begin{aligned}
e_Q=[\kappa(Q):\mathbb F_2],\quad
w_Q=|\operatorname{ord}_Q(\bar x)|+|\operatorname{ord}_Q(\bar y)|+|\operatorname{ord}_Q(\bar z)|,\quad
W_P=\sum_{Q\in S_P}e_Qw_Q.
\end{aligned}
\label{eq:valuation_weights}
\end{equation}
Here $\kappa(Q)$ is the residue field at $Q$. Since $Q$ is a closed point on a curve of finite type over $\mathbb F_2$, Zariski's lemma implies $e_Q<\infty$~\cite{atiyah2018introduction}, and these degrees are uniformly bounded as $Q$ varies over the finite set $S_P$. Since $S_P$ is finite, $W_P$ is a finite constant depending only on the branch $P$.

Let $H\in\mathcal R$ satisfy $\operatorname{Deg}(H)\le d$ and $\eta_P(H)\ne0$. Every monomial in $H$ has the form $x^uy^vz^w$ with $|u|,|v|,|w|\le d$. 
By Eq.~\eqref{eq:discrete_valuation_rules},
\begin{equation}
\operatorname{ord}_Q(\bar x^u\bar y^v\bar z^w)=u\operatorname{ord}_Q(\bar x)+v\operatorname{ord}_Q(\bar y)+w\operatorname{ord}_Q(\bar z)\ge-w_Q d.
\end{equation}
Summing the monomials of $H$ then gives
\begin{equation}
\operatorname{ord}_Q(\bar H)\ge-w_Q d.
\label{eq:polynomial_valuation_lower}
\end{equation}

For an upper bound, we use the fact that a principal divisor on a normal projective curve has weighted degree zero~\cite{liu2002algebraic}:
\begin{equation}
\sum_{Q\ \mathrm{closed}}
[\kappa(Q):\mathbb F_2]\operatorname{ord}_Q(\bar H)=0.
\label{eq:principal_divisor_degree_zero}
\end{equation}
If $Q'\notin S_P$, then $\bar x,\bar y,\bar z$ and their inverses are all regular at $Q'$. 
Thus $\bar H$ has no pole there, so $\operatorname{ord}_{Q'}(\bar H)\ge0$. Fix $Q\in S_P$ and apply Eq.~\eqref{eq:polynomial_valuation_lower} at the other boundary points to obtain
\begin{equation}
e_Q\operatorname{ord}_Q(\bar H)=-\sum_{Q'\ne Q}[\kappa(Q'):\mathbb F_2]\operatorname{ord}_{Q'}(\bar H)
\le d\sum_{Q'\in S_P\setminus\{Q\}}e_{Q'}w_{Q'}=d(W_P-e_Qw_Q).
\label{eq:polynomial_valuation_upper}
\end{equation}
Therefore
\begin{equation}
-dw_Q
\le\operatorname{ord}_Q(\bar H)
\le d\left(\frac{W_P}{e_Q}-w_Q\right).
\label{eq:polynomial_valuation_interval}
\end{equation}

Now set $d=\rho+C_{P,j,i}$ and apply this estimate to $H=A,B$. Both valuations lie in the same interval in Eq.~\eqref{eq:polynomial_valuation_interval}, whose length is $dW_P/e_Q$. Hence
\begin{equation}
\left|\operatorname{ord}_Q\left(\frac{\bar A}{\bar B}\right)\right|=|\operatorname{ord}_Q(\bar A)-\operatorname{ord}_Q(\bar B)|\le d\frac{W_P}{e_Q}\le(\rho+C_{P,j,i})W_P\le(1+C_{P,j,i})W_P\rho,
\end{equation}
where we used $e_Q\ge1$ and $\rho\ge1$. We may therefore take
\begin{equation}
C'_{P,j,i}=(1+C_{P,j,i})W_P.
\end{equation}
This constant is independent of $A,B,\rho$, and $Q$, proving the claim.
\end{proof}


\section{Minimum stabilizer weight theorem and a collection of TB codes supporting PSCQM}
\label{sec:results}

In this section, we present two theorems: the first establishes that the minimum stabilizer weight for any PSCQM satisfying the criterion Eq.~\eqref{eq:criterion} based on
TB codes is $6$, and the second provides a collection of TB codes supporting PSCQM. 

\begin{theorem}
	\label{thm:minimum_tb_weight}
	Every TB code generated by $f,g\in \mathcal R$ satisfying criterion~\eqref{eq:criterion} must have $W\ge6$.
	Here, $W=\operatorname{wt}(f)+\operatorname{wt}(g)$, where $\operatorname{wt}(h)=|\operatorname{supp}(h)|$
	for a Laurent polynomial $h$.
\end{theorem}

\begin{proof}
	Properness of $I=\langle f,g\rangle$ excludes a monomial generator, since every Laurent monomial is a unit.
	In addition, neither generator can vanish: if $f=0$, then $\gcd(f,g)=1$ would force $g$ to be a unit, contradicting properness; the same argument applies to $g$.
	Thus $\operatorname{wt}(f),\operatorname{wt}(g)\ge2$.
	If $W<6$, at least one generator is a binomial, say $f=m_1+m_2=m_1(1+m)$, where $m=m_2/m_1=x^ay^bz^c$ and $\vec v=(a,b,c)^T\ne0$.
	The proper ideal $I$ has a minimal prime $P$.
	Since $f\in P$ and $m_1$ is a unit, $\bar m=1$ in $\mathcal R/P$.
	At every boundary point $Q\in S_P$, additivity of orders gives
	\begin{equation}
		0=a\operatorname{ord}_Q(\bar x)+b\operatorname{ord}_Q(\bar y)+c\operatorname{ord}_Q(\bar z).
	\end{equation}
	Hence $V_P\vec v=0$, contradicting $\operatorname{rank}_{\mathbb Q}V_P=3$.
	Therefore $W\ge6$.
	The lower bound is sharp as we find $f,g$ with $W=6$ satisfying criterion~\eqref{eq:criterion} in the following theorem.  
\end{proof}

\begin{theorem}
	\label{thm:basic_tb_family}
	Consider TB codes generated by 
	\begin{equation}
		f=1+x+y,\quad g=1+x^a+z,\quad a\in\mathbb N^+.
		\label{eq:basic_tb_family}
	\end{equation}
	The polynomials $f,g$ satisfy criterion~\eqref{eq:criterion} so that the corresponding Hamiltonian 
	constitutes a PSCQM if $a$ is not a power of $2$.
\end{theorem}

\begin{proof}
	As the quotient ring $\mathcal R/\langle f,g\rangle$ sets $y=1+x$ and $z=1+x^a$, we have
	\begin{equation}
		\mathcal R/\langle f,g\rangle\cong\mathbb F_2[t^{\pm1},(1+t)^{-1},(1+t^a)^{-1}],
		\label{eq:basic_tb_quotient}
	\end{equation}
	where $x,y,z$ are mapped to $t,1+t,1+t^a$, respectively.
	The polynomials $f,g$ have no nonunit common divisor, so $\gcd(f,g)=1$.
	Moreover, $\mathbb F_2[t^{\pm1},(1+t)^{-1},(1+t^a)^{-1}]$ is a nonzero subring of $\mathbb F_2(t)$ and hence an integral domain.
	Thus $\langle f,g\rangle$ is a proper prime ideal, and $\operatorname{Min}(\langle f,g\rangle)=\{\langle f,g\rangle\}$.
	
	For the unique minimal prime $P=\langle f,g\rangle$, let $h(t)=t(1+t)(1+t^a)$, so that $\mathbb F_2[t^{\pm1},(1+t)^{-1},(1+t^a)^{-1}]=\mathbb F_2[t,h(t)^{-1}]$.
	Thus $\operatorname{Spec}(\mathcal R/\langle f,g\rangle)$ is an open subscheme of the affine line defined by $h(t)\ne0$.
	Consider the standard graded ring
	\begin{equation}
		\mathcal C_P=\mathbb F_2[T,U],\ \deg T=\deg U=1.
		\label{eq:basic_tb_projective_ring}
	\end{equation}
	The associated projective scheme is $\operatorname{Proj}(\mathcal C_P)\cong\mathbb P^1_{\mathbb F_2}$, with homogeneous coordinates $[T:U]$.
	On the chart $U\ne0$, set $t\mapsto T/U$; its coordinate ring is $\mathbb F_2[t]$.
	The above open subscheme is thereby identified with
	\begin{equation}
		U_P=D_+\!\bigl(TU(T+U)(T^a+U^a)\bigr),
	\end{equation}
	where $D_+(F)=\{p\in \mathrm{Proj}(\mathbb F_2[T,U]) \mid F\notin p\}$.
	This open subscheme is nonempty, and the projective line is irreducible, so $U_P$ is dense.
	Hence $\operatorname{Proj}(\mathcal C_P)$ is a projective closure of the original affine curve.
	
	As a polynomial ring over a field, $\mathcal C_P$ is a unique factorization domain and is therefore integrally closed in its fraction field $\mathbb F_2(T,U)$; equivalently, $\mathcal C_P$ itself is normal.
	Therefore,
	\begin{equation}
		\bar{\mathcal C}_P=\mathcal C_P=\mathbb F_2[T,U].
		\label{eq:basic_tb_normalized_ring}
	\end{equation}
	The normalized projective curve is still $\mathbb P^1_{\mathbb F_2}$, with function field $\mathbb F_2(t)$, and
	\begin{equation}
		\bar x=t,\ \bar y=1+t,\ \bar z=1+t^a.
		\label{eq:basic_tb_coordinates}
	\end{equation}
	
	The set of boundary points is $S_P=\operatorname{Proj}(\mathcal C_P)\setminus U_P$.
	Choose the three closed points
	\begin{equation}
		Q_0=[0:1],\quad Q_1=[1:1],\quad Q_\infty=[1:0].
	\end{equation}
	They satisfy $T=0$, $T+U=0$, and $U=0$, respectively, so all three lie on the boundary.
	Factoring the positive integer $a$ into a power of $2$ and an odd factor gives the unique decomposition $a=2^s b$, where $s\in\mathbb N_0$ and $b$ is a positive odd integer.
	The order of a function can be read off by expressing it as an integer power of a local parameter times a unit.
	At $Q_0$, $t$ is a local parameter, whereas $1+t$ and $1+t^a$ are units, because their values at $t=0$ are both $1$.
	Thus the orders of $\bar x,\bar y,\bar z$ at this point are $1,0,0$, respectively.
	At $Q_1$, $1+t$ is a local parameter and $t$ is a unit.
	The identity in characteristic $2$,
	\begin{equation}
		1+t^a=(1+t)^{2^s}h_b(t)^{2^s},\ h_b(t)=\sum_{j=0}^{b-1}t^j,
	\end{equation}
	together with the fact that $b$ is odd and hence $h_b(1)=1$, shows that $h_b$ is a unit at $Q_1$.
	The three orders are therefore $0,1,2^s$, respectively.
	At $Q_\infty$, take the local parameter $u=t^{-1}$. Then
	\begin{equation}
		\bar x=u^{-1},\ \bar y=u^{-1}(1+u),\ \bar z=u^{-a}(1+u^a).
	\end{equation}
	The parenthesized factors are units at $u=0$, so the three orders are $-1,-1,-a$, respectively.
	
	The orders at the three points form a submatrix of $V_P$, 
	\begin{equation}
		M_a=\begin{pmatrix}
			1&0&0\\
			0&1&2^s\\
			-1&-1&-a
		\end{pmatrix},\ \det M_a=2^s-a.
		\label{eq:basic_tb_valuation_minor}
	\end{equation}
	If $a$ is not a power of $2$, then $b>1$ and this minor is nonzero. 
	Hence $\operatorname{rank}V_P=3$, and all three conditions of the criterion hold.
	Otherwise, $a=2^s$ so that $\det M_a=0$ and $\operatorname{rank}V_P<3$.
	In fact, in this case, $\bar z=\bar y^a$, so for every boundary point $Q\in S_P$, we have
	\begin{equation}
		\operatorname{ord}_Q(\bar z)=a\operatorname{ord}_Q(\bar y),
	\end{equation}
	implying that the third column of $V_P$ is $a$ times the second. 
	Thus, $\operatorname{rank}V_P=2$.
\end{proof}


\section{Efficient decoders}
\label{sec:efficient_decoders}

We have established in Sec.~\ref{sec:sufficiency_necessity} that TB code families satisfying Eq.~\eqref{eq:criterion} form PSCQSs.
To obtain a PSCQM, we need to construct a decoder that runs in polynomial time and corrects all low-energy-barrier errors required by the memory-lifetime argument.
We use a renormalization group (RG) decoder~\cite{bravyi2013quantum} and combine its correctability guarantee with the thermal recovery estimate used there.
Throughout, $f,g$ and the energy scale $J$ are fixed.

Our decoder is a specialization of the RG decoder to TB codes.
At each scale, the RG decoder groups the current syndrome defects into clusters according to their separation and applies a local solver to each cluster.
Clusters that can be annihilated locally are removed and their correction operators are recorded; the remaining clusters are carried over to the next scale, at which the clustering distance doubles.
For TB codes, the local correction is obtained by solving a binary linear system.

Starting from $r=1$, we identify each nonzero syndrome entry with a defect.
We connect two defects whenever their $\ell_\infty$ distance is at most $r$ and take each connected component to be a cluster.
Connectivity is transitive: if $A$ is linked to $B$ and $B$ to $C$, all three belong to the same cluster.
For each cluster, we take its smallest enclosing rectangular box and enlarge it by a fixed-width neighborhood to define the local decoding region.
We then restrict the correction qubits to the candidate region and solve the local syndrome equations using the corresponding check matrix $H_X$ or $H_Z$ over $\mathbb F_2$.
Whenever a solution exists, we record the correction operator and remove the corresponding cluster of defects; otherwise, the cluster remains.
If defects remain, we set $r\leftarrow2r$ and repeat the clustering and local decoding steps.
Once the syndrome is empty, the decoder returns the accumulated correction; if defects remain after the prescribed maximum number of levels, it reports failure. 
Denote this classical syndrome-to-correction map by $D_{\mathrm{RG}}$; its successful outputs are Pauli operators. 
The associated quantum recovery channel is the map $\mathcal L_L^{\mathrm{dec}}$ defined in Sec.~\ref{sec:psc_codes}.

Subsection~\ref{sec:sufficiency} shows that the TB code families considered here satisfy TQO1, TQO2, and UNSC.
Reference~\cite{bravyi2013quantum} guarantees that the RG decoder correctly recovers from any error $E$ satisfying $(80\alpha_*)^{\Delta(E)/(2J)}<T_*$, where $\alpha_*$ is the uniform no-strings constant and $T_*=\Theta(L)$ is the applicable topological scale in the conventions of that reference.
Consequently, there are constants $a>0$ and $b\in\mathbb R$, independent of $L$, such that, for all sufficiently large $L$,
\begin{equation}
	\Delta(E)<a\log n_L+b\Longrightarrow D_{\mathrm{RG}}\circ\operatorname{syn}(E)E\in\mathcal S_L
	\label{eq:rg_decoder_criterion}
\end{equation}
holds for every Pauli error $E$, with overall phases ignored.
The condition on the right guarantees recovery of any initial encoded state.

Polynomial runtime follows by counting the local elimination steps.
Each level contains at most $O(n_L)$ clusters, and each local linear system has at most $O(n_L)$ rows and columns, so binary Gaussian elimination takes $O(n_L^3)$ time per system.
There are $O(\log n_L)$ levels, giving time $O(n_L^4\log n_L)$.

Equation~\eqref{eq:rg_decoder_criterion} shows that every error uncorrectable by the RG decoder has an energy barrier of at least $a\log n_L+b$.
The decoder and code family can therefore be viewed together as an efficiently decodable system with an effective energy barrier bounded below by $a\log n_L+b$.


\section{Numerical simulation}

\label{sec:numerical_results}

In this section, we describe the numerical method and simulation settings used for Fig.~3 in the main text.

We simulate the Lindblad evolution under the local Pauli Davies bath defined above, with $\gamma_0=1$ and time measured in units of $\gamma_0^{-1}$.
We set $J=1/2$ in Eq.~\eqref{eq:stabilizer_hamiltonian} and $k_{\mathrm B}=1$, so each stabilizer violation costs one unit of energy.
For a stabilizer Hamiltonian and an initial state $\rho_{0}$ in the code space, this evolution can be sampled using the Gillespie algorithm~\cite{gillespie1977exact,gillespie1976general}.
Each trajectory is represented by its accumulated Pauli operator, or Pauli frame, $P_t$, initialized as $P_0=I$.
Its conditional state is $P_t\rho_{0}P_t^\dagger$, and averaging these states over trajectories gives the solution of the Davies master equation~\cite{bravyi2013quantum,molmer1993monte}.
The stabilizer generators that anticommute with $P_t$ determine the current syndrome and excitation energy.

At the current time $t$, we consider all $2n_L$ single-qubit channels $\alpha=(q,\sigma)$, with $\sigma\in\{X,Z\}$.
Let $\Delta E_\alpha$ be the system's energy change if $\sigma_q$ is applied to the current configuration.
The channel rates and their total are
\begin{equation}
\gamma_\alpha=\gamma(-\Delta E_\alpha,\beta)=\frac{1}{1+e^{\beta\Delta E_\alpha}},
\quad R=\sum_\alpha\gamma_\alpha,
\label{eq:numerical_glauber}
\end{equation}
where the minus sign converts the system's energy change to the energy released to the bath in Eq.~\eqref{eq:davies_rate}.
We draw a uniform random number $u\in(0,1)$ and advance to the next jump at
\begin{equation}
\Delta t=-\frac{\ln u}{R}.
\label{eq:numerical_waiting_time}
\end{equation}
Independently, we select a channel $\alpha$ with probability $\gamma_\alpha/R$, apply its Pauli operator, and update
$t\leftarrow t+\Delta t$ and $P_t\leftarrow\sigma_qP_t$.
We then update the syndrome and channel rates and repeat.
This procedure samples the continuous-time thermal trajectory without a fixed time step.

Along each trajectory, we use a decoder to test whether the accumulated error can be recovered.
The proposed correction $C$ is used only for this test and is not applied to the evolving Pauli frame.
Recovery succeeds if the decoder converges, the correction reproduces the input syndrome, and $CP_t\in\mathcal S_L$ up to a global phase.
A failure in either CSS sector is counted as a readout failure of the full logical register.
We record the first detected failure time $T_i$ for each trajectory and estimate the mean lifetime $\tau_{L}(\beta)=\mathbb E[T_i]$.
For a complete sample of $N$ independent trajectories, the sample mean and its standard error are
\begin{equation}
\widehat\tau_{a,L}(\beta)=\frac{1}{N}\sum_{i=1}^{N}T_i,\quad\operatorname{SE}(\widehat\tau_{a,L})=\sqrt{\frac{\sum_{i=1}^{N}(T_i-\widehat\tau_{a,L})^2}{N(N-1)}}.
\label{eq:numerical_mean_estimator}
\end{equation}

We simulate the two TB code families with $f=1+x+y$ and $g=1+xy+z$ or $g=x+z+z^{-1}$, using a target sample of $N=400$ independent trajectories for each $(g,L,\beta)$.
The inverse temperatures are $\beta=4,4.25,\ldots,6.5$.
The scanned periods are $L\in\{7,21,35,49,77,91\}$ for $g=1+xy+z$ and $g=x+z+z^{-1}$.
Within each family, the periods are chosen to keep the number of logical qubits fixed throughout the size scan: $k_L=12$ for for both families.
Thus, the lifetime comparison across sizes concerns a logical register of fixed size within each family.

Readout uses the sparse-matrix belief-propagation (BP) decoder in \texttt{ldpc}~2.4.0~\cite{roffe2022ldpc,roffe2020decoding}, applied separately to the two CSS sectors~\cite{yedidia2005constructing}.
We use minimum-sum message updates, a fixed serial update order, at most 100 iterations, a message scaling factor of 1, and a fixed independent-error prior $p_{\rm BP}=0.05$.
These decoder settings are held constant across all sizes and temperatures.
\end{widetext}

\end{document}